\documentclass[aps,pra,showpacs,preprint]{revtex4-1}
\usepackage{epsfig,graphicx,color}

\begin{document}

\title{Bright solitons in quasi-one dimensional dipolar condensates with
spatially modulated interactions}
\author{F.Kh. Abdullaev$^{1}$}
\author{A. Gammal$^{2}$}
\author{B. A. Malomed$^{3}$}
\author{Lauro Tomio $^{1,4}$}
\affiliation{$^1$Instituto de F\'\i sica Te\'orica, Universidade Estadual Paulista,
01140-070, S\~ao Paulo, SP, Brazil\\
$^2$Instituto de F\'\i sica, Universidade de S\~ao Paulo, 05508-090, S\~ao
Paulo, SP, Brazil\\
$^3$Department of Physical Electronics, School of Electrical Engineering,
Faculty of Engineering, Tel Aviv University, Tel Aviv 69978, Israel\\
$^4$ Centro de Ci\^encias Naturais e Humanas, Universidade Federal do ABC,
09210-170, Santo Andr\'e, Brazil}
\date{\today }

\begin{abstract}
We introduce a model for the condensate of dipolar atoms or molecules, in
which the dipole-dipole interaction (DDI) is periodically modulated in
space, due to a periodic change of the local orientation of the permanent
dipoles, imposed by the corresponding structure of an external field (the
necessary field can be created, in particular, by means of magnetic
lattices, which are available to the experiment). The system represents a
realization of a nonlocal nonlinear lattice, which has a potential to
support various spatial modes. By means of numerical methods and variational
approximation (VA), we construct bright one-dimensional solitons in this
system, and study their stability. In most cases, the VA provides good
accuracy, and correctly predicts the stability by means of the
Vakhitov-Kolokolov (VK)\ criterion. It is found that the periodic modulation
may destroy some solitons, which exist in the usual setting with unmodulated
DDI, and can create stable solitons in other cases, not verified in the
absence of modulations. Unstable solitons typically transform into
persistent localized breathers. The solitons are often mobile, with
inelastic collisions between them leading to oscillating localized modes.
\end{abstract}

\maketitle



\section{Introduction}

Studies of Bose-Einstein condensates (BEC) in optical lattices (OLs) is one
of central topics in the current work on cold atoms \cite{OL,Oxford}. Two
types of spatially-periodic modulation of parameters of BEC are possible
under the action of OLs. First is a periodic (lattice) potential for atoms,
which represents the linear OL. The spatial periodicity of another kind can
be realized in BEC via a modulation of the atomic scattering length, i.e.,
the effective local nonlinearity of the condensate, which is known as the
nonlinear OL (or a magnetic lattice \cite{ML}, if the scattering length is
affected by the magnetic field). New types of solitons and solitary vortices
are possible in nonlinear lattices \cite{KMT}.

Both linear and nonlinear OLs can be realized in condensates with atoms
interacting via long-range dipolar-dipolar interactions (DDI). Due to the
anisotropic and nonlocal character of the DDI, many remarkable new phenomena
have been predicted and observed in these ultracold gases, such as new
quantum phases, anisotropic collapse and suppression of the collapse, and
specific modes of collective excitations \cite{Baranov,Lahaye}, as well as
stabilization of modulated patterns trapped in the OL potential, which are
unstable in the BEC with the contact interaction \cite{Pfau}. The DDI may
also give rise to localized modes in the form of stable two-dimensional 
(2D)~\cite{2D} and one-dimensional (1D)~\cite{JCuevas,Luis} solitons and soliton
complexes \cite{KLakomy}. In fact, the presence of the linear OL potential
strongly facilitates the creation and stabilization of solitons in dipolar
condensates \cite{JCuevas,Belgrade,SKA,RVMA}.

A new type of lattice structures can be created in dipolar condensates, by
means of a spatially periodic modulation of the strength of the DDI. In
fact, these are \textit{nonlocal nonlinear lattices}, as the DDI represents
long-range cubic interactions. Nonlocal nonlinear lattices of different
types are possible too in optical media with the thermal nonlinearity~\cite
{Kartashov}. In the dipolar BEC, one possibility for the creation of such a
lattice is to use a condensate of atoms or molecules with permanent magnetic
or electric dipolar moments, and apply a spatially nonuniform (misaligned)
polarizing field, under the action of which the mutual orientation of the
moments varies in space periodically. Another approach may make use of a
condensate of polarizable atoms or molecules, in which the moments are
induced by the external field \cite{polarizable}, whose strength varies
periodically along the system. In either case, the necessary periodically
nonuniform external magnetic fields can be induced by the currently
available magnetic lattices~\cite{ML}. If the condensate is formed by
particles with permanent or induced electric moments, the corresponding
electric field can be supplied by similarly designed ferroelectric lattices.
The electrostriction effects on DDI induced by a far-off-resonant standing
laser field can also lead to a periodic structure in the dipolar BEC~\cite
{Gershon}. Such systems are described by the Gross-Pitaevskii (GP) equation
with a spatially varying nonlocal nonlinearity.

The objective of the present work is to investigate the existence and
stability of effectively 1D~\cite{Roati,Santos} solitons in the dipolar BEC
under the action of nonlocal nonlinear lattices. Here we focus on the model
which assumes permanent moments with a periodically varying orientation. The
soliton modes will be studied by means of a variational approximation (VA),
and with the help of direct numerical simulations of the respective GP
equation. For direct simulations, we have employed a relaxation algorithm
applied to the GP equation, as described in Ref.~\cite{BGT}.

The paper is organized as follows. The model is introduced in Section II.
Analytical results (the VA, and some simplification of the full model) are
reported in Section III. Numerical results are summarized and compared to
the variational predictions in Section IV, and the paper is concluded by
Section V.

\section{The model}

Under the assumption that that the external field imposes an orientation 
of permanent magnetic moments periodically varying along the $x-$axis, 
the corresponding dipolar BEC obeys the following nonlocal GP equation\cite{Santos,JCuevas,YCai}:
\begin{equation}
\mathrm{i}\hbar \frac{\partial \Psi }{\partial {t}}+\frac{\hbar ^{2}}{2m}
\frac{\partial ^{2}\Psi }{\partial {x^{2}}}-2a_{s}^{(r)}\hbar \omega _{\perp
}|\Psi |^{2}\Psi -\frac{2d^{2}}{l_{\perp }^{3}}\Psi \int_{-\infty }^{+\infty
}d\xi \;f(x,\xi )R(|x-\xi |)\;|\Psi (\xi )|^{2}=0.  \label{eq01}
\end{equation}
Here, $a_{s}^{(r)}$ is the renormalized $s$-wave scattering length, which
includes a contribution from the DDI:
\begin{equation}
a_{s}^{(r)}=a_{s}\left[ 1+\frac{\epsilon _{\mathrm{DD}}}{2}(1-3\cos
^{2}\theta _{0})\right] ,\;\;\;\mathrm{with}\;\;\;\epsilon _{\mathrm{DD}}=
\frac{md^{2}}{3\hbar ^{2}a_{s}}\equiv \frac{a_{d}}{3a_{s}},  \label{eq02}
\end{equation}
where $a_{s}$ is the proper scattering length, which characterizes
collisions between the particles, $\theta _{0}$ is the average value of the
dipole's orientation angle, and $a_{d}$, defined by the dipole momentum $d$
and the atomic mass $m$, plays the role of the effective DDI scattering
length. In the above expressions, $\omega _{\perp }$ is the transverse
trapping frequency, which defines the respective trapping radius, $l_{\perp
}=\sqrt{\hbar /m\omega _{\perp }}$. The reduction of the full GP
equation to its 1D form (\ref{eq02}) is valid provided that the peak density
of the condensate, $n_{0}$, is small enough to make the nonlinear
term much weaker than the transverse confinement, and prevent excitation of
transverse models by the DDIs, i.e.,
\[
\int d^{3}r^{\prime }U_{dd}(\vec{r}-\vec{r}^{\prime })|\Phi ({\vec{r}}
^{\prime },t)|^{2}\ll \hbar \omega _{\perp },\;\;U_{dd}(\vec{r}-\vec{r}
^{\prime })=\frac{d^{2}(1-3\cos ^{2}(\theta _{0}))}{4\pi \epsilon _{0}|\vec{r
}-\vec{r}^{\prime }|^{3}},
\]
where $\epsilon _{0}$ is the vacuum permittivity. In this case, the
3D wave function may be factorized as $\Phi ({\vec{r}},t)\approx \Psi (x,t)
\mathcal{R}(\rho )$, with the transverse radial mode $\mathcal{R}(\rho )$
taken as the ground state of the confining harmonic-oscillator
potential. After performing the integration over the transverse radius 
$\rho $ in the GP equation, we arrive at the quasi-1D equation (\ref{eq01}).

The exact DDI\ kernel $R(x)$ appearing in Eq. (\ref{eq01}) has a rather
cumbersome form \cite{Santos,YCai}, but it may be approximated reasonably
well by~\cite{JCuevas}
\begin{equation}
R(x)=\frac{\epsilon ^{3}}{(x^{2}+\epsilon ^{2})^{3/2}},  \label{eq04}
\end{equation}
where a cutoff parameter $\epsilon $, which truncates the formal
singularity, is on the order of the transverse radius of the cigar-shaped
trap, $l_{\perp }$. In the following, the variables are rescaled as
\begin{equation}
x\rightarrow {l_{\perp }}x,\;\;\xi \rightarrow {l_{\perp }}x^{\prime
},\;\;t\rightarrow t/\omega _{\perp },\;\;2a_{s}^{r}/l_{\perp }\rightarrow
g,~\Psi (x,t)\rightarrow \psi (x,t)/\sqrt{l_{\perp }},  \label{rescale}
\end{equation}
with the corresponding dimensionless cutoff parameter fixed to be $\epsilon
=1$.

Further, the function $f(x,\xi )$ in Eq.~(\ref{eq01}) describes the
modulation of the DDI due to the variation of the dipoles' orientation along
the $x-$axis,
\begin{equation}
f(x,\xi )=\cos \left( \theta (x)-\theta (\xi )\right) -3\cos \left( \theta
(x)\right) \cos \left( \theta (\xi )\right) ,  \label{eq03}
\end{equation}
where we assume that the local angle of the orientation of the dipoles with
respect to the $x-$axis is periodically modulated as follows:
\begin{equation}
\theta (x)=\theta _{0}+\theta _{1}\cos \left( {kx}\right) .  \label{eq07}
\end{equation}
The periodic variation of $\theta (x)$ can be induced by magnetic
lattices (MLs), which have been constructed on the basis of ferromagnetic
films with permanent periodic structures built into them, the condensate
being loaded into the ML \cite{ML}. Accordingly, the ground state of the
dipolar BEC is formed by an elongated trap with a strong magnetic field $B$,
whose relevant value may be estimated as $\simeq $200 G.
The periodic modulation of the DDI is then induced by the ML with the
amplitude of the respective component of the magnetic field $B\sim 40-50$ G
and a modulation period $\lambda \sim 0.5$ $\mathrm{\mu }$m, which is within
the same order of the magnitude as a typical OL period. The regular ML can
be constructed to include $\sim 1000$ sites, which is quite sufficient to
observe the localized modes described below~\cite{ML}.

With regard to the rescaling (\ref{rescale}), the combined kernel in 
Eq.~(\ref{eq01}) is transformed as 
$2d^{2}/(l_{\perp }^{3}\hbar \omega _{\perp
})\;f(x,\xi )R(|x-\xi |)\rightarrow G\,V_{\mathrm{DD}}(x,x^{\prime })$,
where $G\equiv 2d^{2}/(l_{\perp }^{3}\hbar \omega _{\perp })$ is the DDI
effective strength, and
\begin{eqnarray}
V_{\mathrm{DD}}(x,x^{\prime }) &=&\frac{\cos \left( \theta (x)-\theta \left(
x^{\prime }\right) \right) -3\cos \left( \theta (x)\right) \cos \left(
\theta \left( x^{\prime }\right) \right) }{\left[ \left( x-x^{\prime
}\right) ^{2}+\epsilon ^{2}\right] ^{3/2}}  \nonumber \\
&\equiv &-\;\frac{\cos \left( \theta (x)-\theta \left( x^{\prime }\right)
\right) +3\cos \left( \theta (x)+\theta \left( x^{\prime }\right) \right) }{
2 \left[ \left( x-x^{\prime }\right) ^{2}+\epsilon ^{2}\right] ^{3/2}}.
\label{eq05}
\end{eqnarray}
Thus, the underlying GP equation is cast into the form of the following
partial-integral differential equation, which combines the contact and
dipole-dipole interactions:
\begin{equation}
\mathrm{i}\frac{\partial \psi (x)}{\partial t}=-\frac{1}{2}\frac{\partial
^{2}\psi (x)}{\partial x^{2}}+g|\psi (x)|^{2}\psi (x)+G\psi (x)\int_{-\infty
}^{+\infty }V_{\mathrm{DD}}(x,x^{\prime })\left\vert \psi \left( x^{\prime
}\right) \right\vert ^{2}dx^{\prime }.  \label{eq06}
\end{equation}

\begin{figure}[tbp]
\centerline{
\includegraphics[width=8cm,clip]{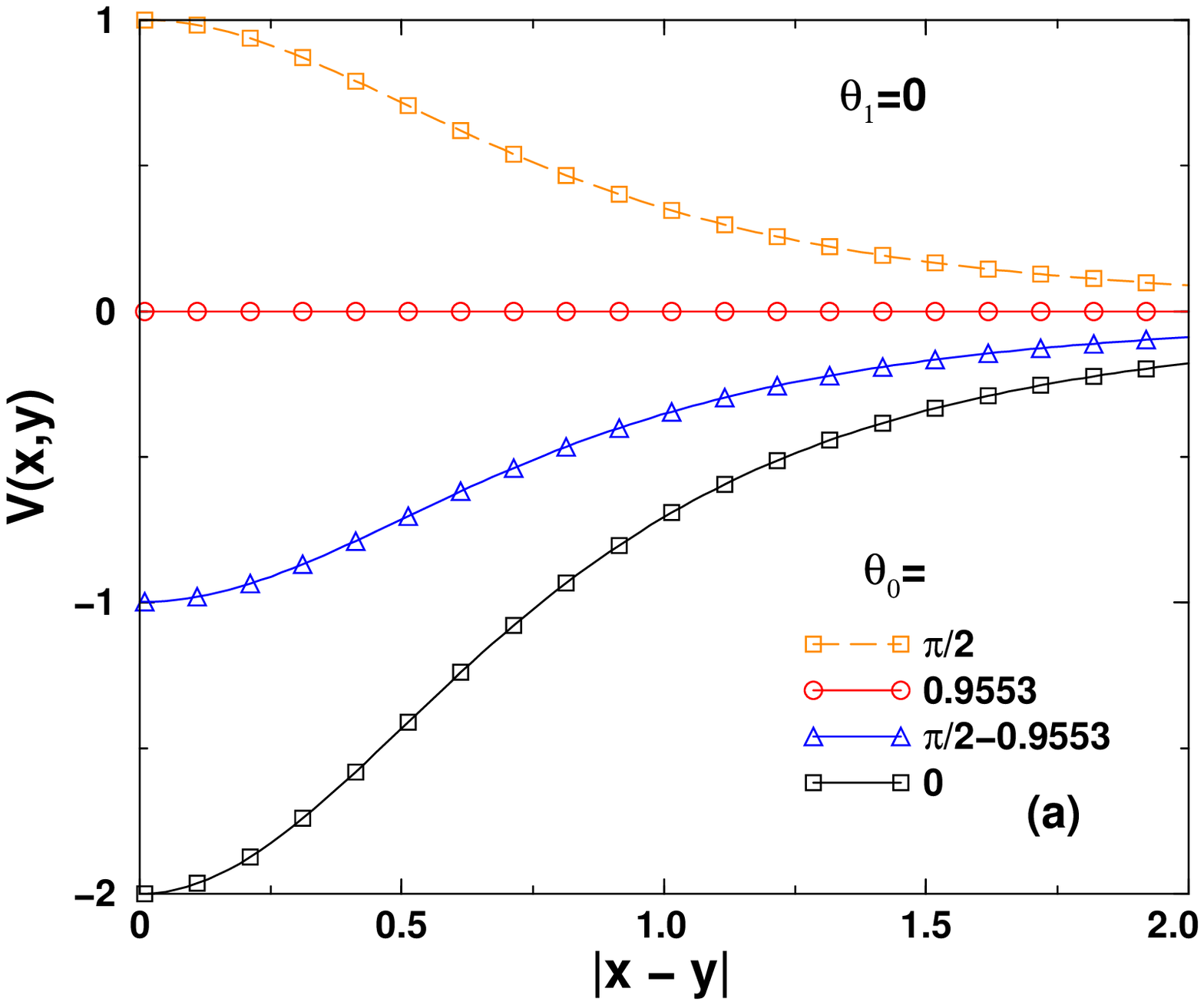}
\includegraphics[width=8cm,clip]{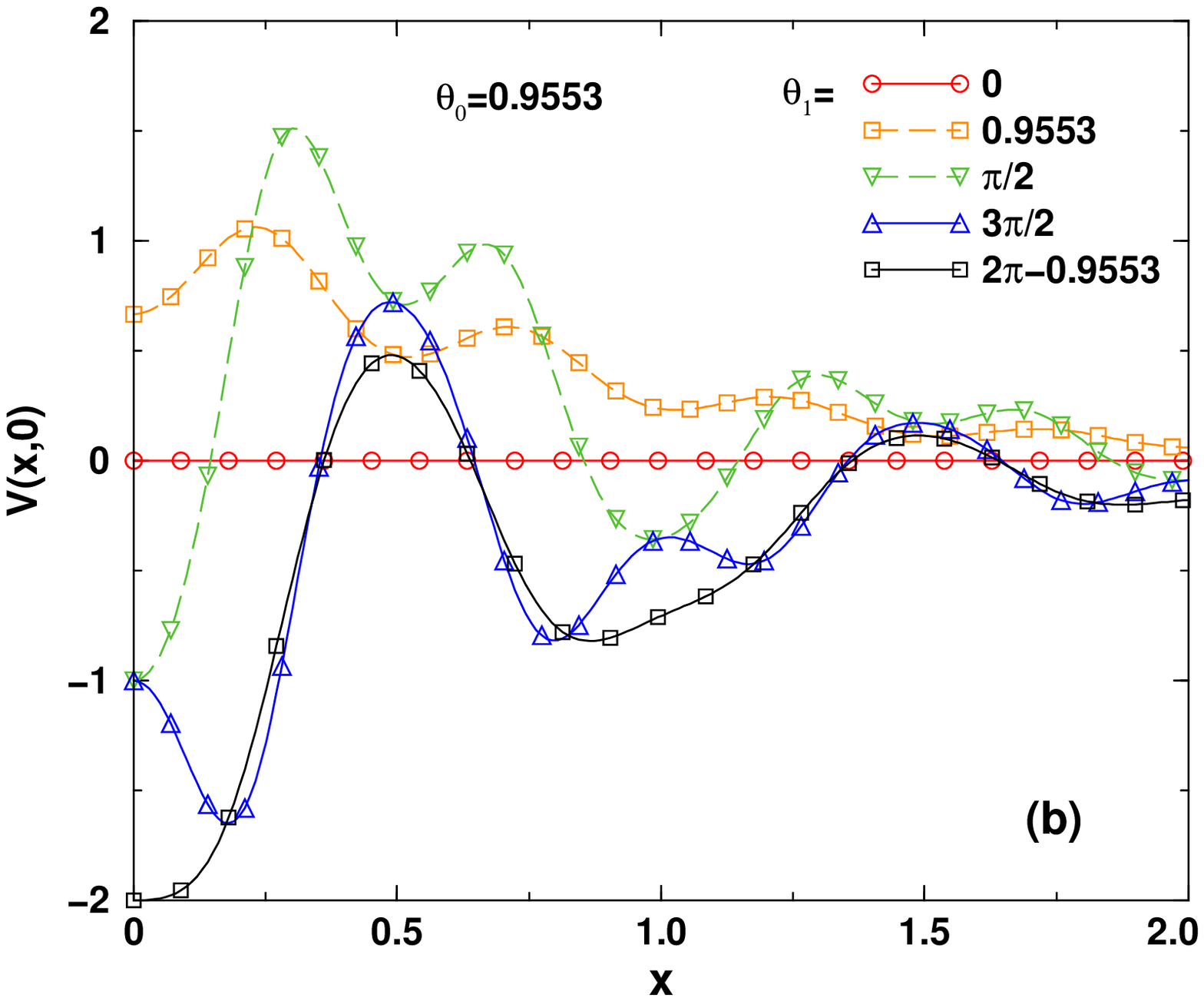}
}
\centerline{
\includegraphics[width=8cm,clip]{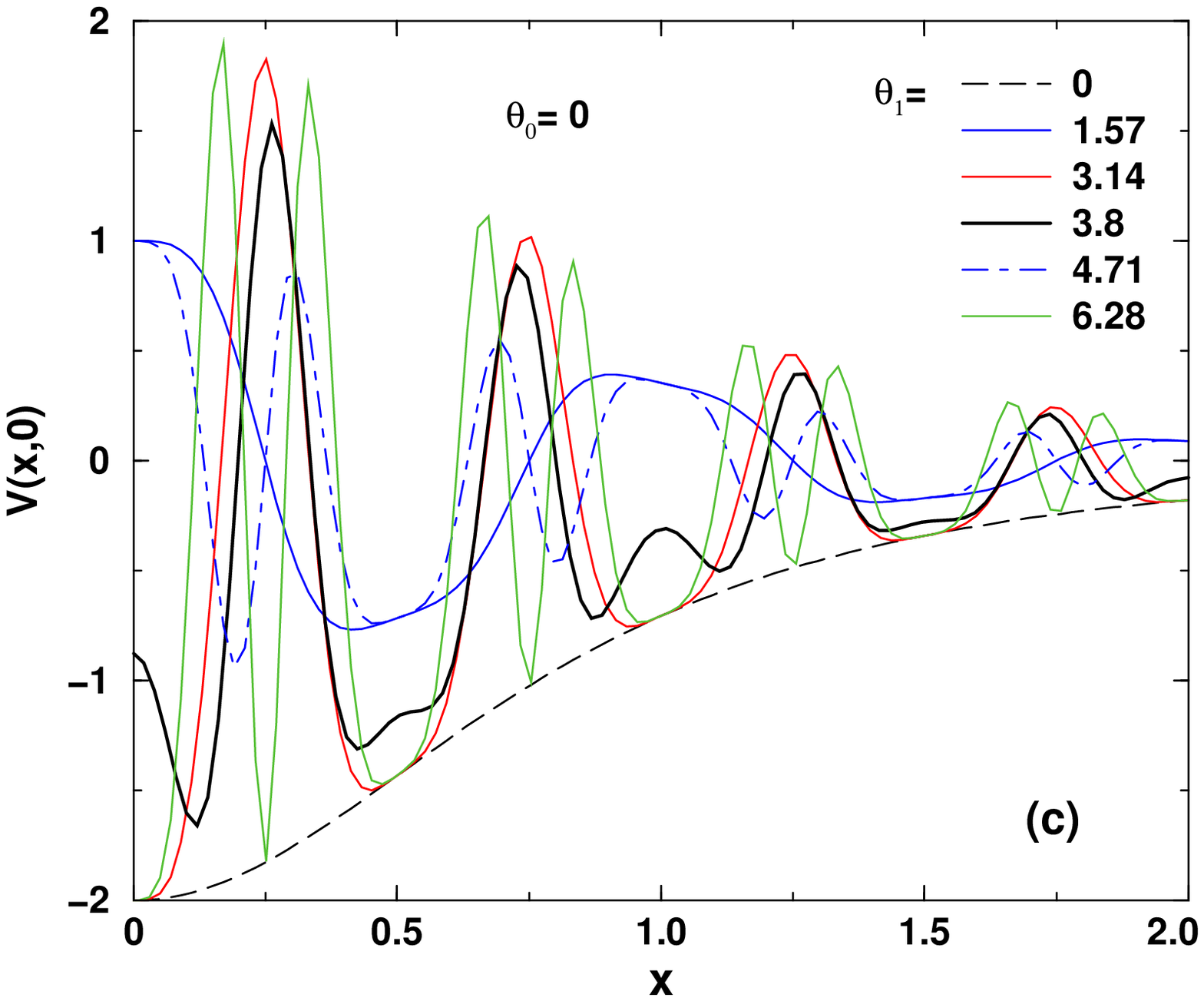}
\includegraphics[width=8cm,clip]{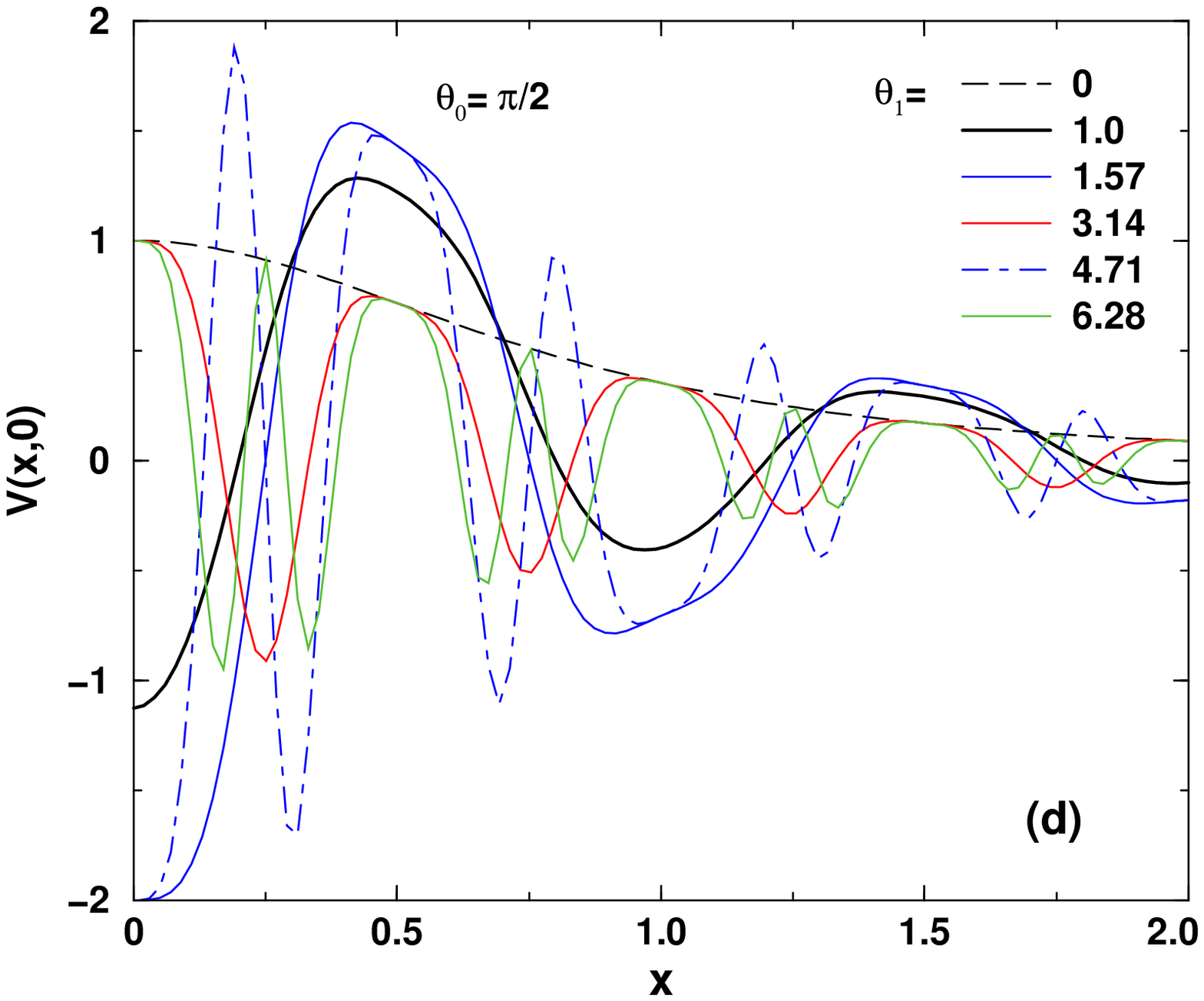}
}
\caption{(Color online) The effective DDI kernel, as given by Eqs.~(\protect
\ref{eq05}) and (\protect\ref{eq07}). The kernel for $\protect\theta _{0}=
\protect\theta _{m}=0.9553$ (in the upper panels) corresponds to the more
specific expression (\protect\ref{eq10}), and the ones for $\protect\theta 
_{0}=0$ and $\protect\theta =\protect\pi /2$ (in the lower panels)
correspond to Eq. (\protect\ref{eq08}). In view of the symmetry of $
V(x,y)$, we choose to fix $y=0$ and take $x>0$, except in the case of the
unmodulated DDI (a), where $V(x,y)$ is given as a function of $|x-y|$.}
\label{Fig-01}
\end{figure}

The effective kernel (\ref{eq05}) can be further simplified for specific
settings. As an example, for the average orientation angles $\theta _{0}=\pi
n/2$, with $n=0$ or $n=1$ (the mean orientations parallel or perpendicular
to the axis, respectively), we have
\begin{equation}
V_{\mathrm{DD}}^{\left( \theta _{0}=\pi n/2\right) }(x,x^{\prime })=-\;\frac{
\cos \left[ \theta _{1}\left( \cos (kx)-\cos (kx^{\prime })\right) \right]
+3(-1)^{n}\cos \left[ \theta _{1}\left( \cos (kx)+\cos (kx^{\prime })\right) 
\right] }{2\left[ \left( x-x^{\prime }\right) ^{2}+\epsilon ^{2}\right]
^{3/2}}.  \label{eq08}
\end{equation}
Another point of interest is the one when the interaction reduces to zero
for $\theta =\theta _{0}$. This happens at
\begin{equation}
\theta _{0}\equiv \theta _{m}\equiv \arccos \left( 1/\sqrt{3}\right) \approx
0.9553.  \label{eq09}
\end{equation}
with the respective kernel being
\begin{equation}
V_{\mathrm{DD}}^{\left( \theta _{0}=\theta _{m}\right) }(x,x^{\prime
})\equiv -\;\frac{\sin \left[ \theta _{1}\cos (kx)\right] \sin \left[ \theta
_{1}\cos (kx^{\prime })\right] -\sqrt{2}\sin \left[ \theta _{1}\left( \cos
(kx)+\cos (kx^{\prime })\right) \right] }{\left[ \left( x-x^{\prime }\right)
^{2}+\epsilon ^{2}\right] ^{3/2}}.  \label{eq10}
\end{equation}
Characteristic shapes of the kernel for four different configurations,
defined by Eqs.~(\ref{eq08}) and (\ref{eq10}), are displayed in Fig.~\ref
{Fig-01}. We choose here some specific configurations of interest.
First, in panel (a) we show the unmodulated case ($\theta _{1}=0$),
ranging from the most attractive ($\theta _{0}=0$) to the most repulsive ($
\theta _{0}=\pi /2$) conditions. In this case, the maximum of $|V(x,y)|$ is
reached when $x=y$. In view of the symmetry of $V(x,y)$, only positive
values for $(x-y)$ are presented. In the other three panels, we fix
$y=0$. In panel (b) we show how this interaction is varying with 
$\theta _{1}$, by fixing $\theta _{0}=\arccos (1/\sqrt{3})$, so that
the interaction vanishes [as also shown in panel (a)] at $\theta _{1}=0$.
In the lower two panels, we plot the effective DDI kernel for the two cases
of interest, by varying $\theta _{1}$ with $\theta _{0}=0$ 
[panel (c)] and $\theta _{0}=\pi /2$ [panel (d)]. 

Thus, the effective (pseudo)potential of the nonlinear interactions in Eq.~(
\ref{eq06}) is
\begin{equation}
V_{\mathrm{eff}}(x;|\psi |^{2})=g\left\vert \psi (x)\right\vert
^{2}+G\int_{-\infty }^{+\infty }V_{\mathrm{DD}}(x,x^{\prime })\left\vert
\psi \left( x^{\prime }\right) \right\vert ^{2}dx^{\prime }.  \label{eq11}
\end{equation}
Coefficients $G$, $g$, $\epsilon $ and $k$ in Eq. (\ref{eq06}) can be easily
rescaled. In view of that, in the following we set $G\equiv 1$, $\epsilon
\equiv 1$, and also choose $k=2\pi $, which fixes the respective period in
Eq. (\ref{eq07}) as $2\pi /k=1$. As a starting point, we set $g\equiv 0$ (no
effective local interaction). Then, the remaining parameters are
coefficients $\theta _{0}$ and $\theta _{1}$ of periodic modulation 
(\ref{eq07}), and the total norm of the solution,
\begin{equation}
N=\int_{-\infty }^{+\infty }\left\vert \psi (x)\right\vert ^{2}dx.
\label{eq12}
\end{equation}
The objective is to find a family of bright-soliton solutions in this model,
and analyze their stability. In the system with the unmodulated DDI, which
also included the linear OL potential, this problem was considered in Ref.
\cite{JCuevas}. A related model (the Tonks-Girardeau gas with the attractive
DDI) was considered in Ref. \cite{BBB}.

Bright solitons may exist if the interaction is self-attractive, i.e., the
sign of the averaged value of the (pseudo)potential (\ref{eq11}) is
negative. We start from the case of $\theta _{1}=0$ (or $k=0$, with $\theta
_{0}$ redefined as $\theta _{1}+\theta _{0}$), when the periodic modulation
is absent. The respective region of values of $\theta _{0}$, where solitons
are likely to occur, can be readily identified. For this case, the
interaction kernel $V_{\mathrm{DD}}(x,y)$ is shown in the panel (a)
of Fig.~\ref{Fig-01}. The corresponding numerator of the expression 
(\ref{eq05}) is $(1-3\cos ^{2}\theta _{0})$, leading to attractive interactions
for all values of $\theta _{0}$ with $\cos ^{2}(\theta _{0})\geq 1/3$.
Within the period of $0<\theta _{0}<2\pi $, this condition holds for 
$0<\theta _{0}<\theta _{m}$, $\pi -\theta _{m}<\theta _{0}<\pi +\theta _{m}$
and $2\pi -\theta _{m}<\theta _{0}<2\pi $ [recall $\theta _{m}$ is given by
Eq.~(\ref{eq09})]. Next, by switching on $\theta _{1}$, the interaction
kernel is made undulating, and the existence of solitons turns out to be
possible outside of these intervals, while in some areas inside the
intervals the solitons do not exist, as we demonstrate below.

\subsection{An approximate analytical approach}

A simplified version of the model can be introduced by using the well-known
Fourier expansions which generate the Bessel functions,
\begin{eqnarray}
\cos (z\cos (\eta )) &=&J_{0}(z)+2\sum_{n=1}^{\infty }(-1)^{n}J_{2n}(z)\cos
(2n\eta ),  \nonumber \\
\sin (z\cos (\eta )) &=&2\sum_{n=0}^{\infty }(-1)^{n}J_{2n+1}\cos
((2n+1)\eta ).  \label{eq13}
\end{eqnarray}
By keeping the lowest-order harmonic in the expansions, we arrive at the
following approximation for interaction kernel (\ref{eq05}) with the \
modulation format (\ref{eq07}):
\begin{eqnarray}
V_{\mathrm{DD}}(x,x^{\prime }) &\approx &\{\sin ^{2}{\theta _{0}}\left[
J_{0}^{2}(\theta _{1})-8J_{1}^{2}(\theta _{1})\cos (kx)\cos (kx^{\prime }) %
\right] +  \nonumber \\
&&3\sin (2\theta _{0})J_{0}(\theta _{1})J_{1}(\theta _{1})[\cos (kx)+\cos
(kx^{\prime })]+  \nonumber \\
&&2\cos ^{2}(\theta _{0})[2J_{1}^{2}(\theta _{1})\cos (kx)\cos (kx^{\prime
})-J_{0}^{2}(\theta _{1})]\}[(x-x^{\prime })^{2}+\epsilon ^{2}]^{-3/2}.
\label{eq14}
\end{eqnarray}
In the particular cases of the dipoles oriented parallel ($\theta _{0}=0$)
and perpendicular ($\theta _{0}=\pi /2$) to the $x-$axis, the expression (%
\ref{eq14}) reduces, respectively, to
\begin{eqnarray}
V_{\mathrm{DD}}^{(\theta _{0}=0)}(x,x^{\prime }) &\approx
&2[2J_{1}^{2}(\theta _{1})\cos (kx)\cos (kx^{\prime })-J_{0}^{2}(\theta
_{1})][(x-x^{\prime })^{2}+\epsilon ^{2}]^{-3/2},  \label{eq15} \\
V_{\mathrm{DD}}^{(\theta _{0}=\pi /2)}(x,x^{\prime }) &\approx
&[J_{0}^{2}(\theta _{1})-8J_{1}^{2}(\theta _{1})\cos (kx)\cos (kx^{\prime
}))][(x-x^{\prime })^{2}+\epsilon ^{2}]^{-3/2}.  \label{eq16}
\end{eqnarray}

The existence condition can be found in an explicit form for broad solitons,
whose width is much larger than that of the kernel (i.e., the GP equation
becomes a quasi-local one). In this case, $|\psi |^{2}$ appearing in the
integral of the nonlocal term of the GP equation may be replaced by a
constant, hence the necessary condition for the existence of a bright
soliton amounts to
\begin{equation}
\int_{-\infty }^{+\infty }V_{\mathrm{DD}}(0,x^{\prime })dx^{\prime }<0.
\label{eq17}
\end{equation}
For the analysis of this condition, we take into account that
\begin{eqnarray}
&&\int_{-\infty }^{+\infty }\frac{1}{\left[ (x^{\prime })^{2}+\epsilon ^{2} %
\right] ^{3/2}}dx^{\prime }=\frac{2}{\epsilon ^{2}},  \nonumber \\
&&\int_{-\infty }^{+\infty }\frac{\cos (kx^{\prime })}{\left[ (x^{\prime
})^{2}+\epsilon ^{2}\right] ^{3/2}}dx^{\prime }=\frac{2|k\epsilon |}{
\epsilon ^{2}}K_{1}(|k\epsilon |),  \label{eq18}
\end{eqnarray}
where $K_{1}(k\epsilon )$ is the modified Bessel functions. Making use of
the symmetry properties of Eqs.(\ref{eq07}) and (\ref{eq05}) and the
periodicity, in the following we assume that $\theta _{0}$ and $\theta _{1}$
are positive. To identify existence regions for broad solitons, we consider
three illustrative cases:

\begin{itemize}
\item[1)] For the dipoles oriented along the $x\ $axis, i.e., with $\theta
_{0}=0$, intervals of $\theta _{1}$ for the existence of the bright soliton
follow from Eqs. (\ref{eq15}) and (\ref{eq17}) in the form of
\begin{equation}
2|k\epsilon |\;K_{1}(|k\epsilon |)\;J_{1}^{2}(\theta _{1})-J_{0}^{2}(\theta
_{1})<0.  \label{eq19}
\end{equation}
Taking, as said above, $\epsilon =1$ and $k=2\pi $, we have $2|k\epsilon
|\;K_{1}(|k\epsilon |)=0.0124$, hence the first three existence intervals
are
\begin{equation}
0<\theta _{1}<2.29,\;\;\;2.52<\theta _{1}<5.40,\;\;\;5.63<\theta
_{1}<8.54.\;\;  \label{eq20}
\end{equation}
In each interval, the maximum strength of the attractive DDI is attained,
severally, at $\theta _{1}=0$, $3.83$, and $7.0$. Virtually the same
existence intervals are produced by numerical solutions (see Section IV
below).

\item[2)] For dipoles oriented perpendicular to the $x\ $axis, condition ( %
\ref{eq19}) is replaced by
\begin{equation}
J_{0}^{2}(\theta _{1})-8|k\epsilon |\;K_{1}(|k\epsilon |)\;J_{1}^{2}(\theta
_{1})<0,  \label{eq21}
\end{equation}
so that, for $\epsilon =1$ and $k=2\pi $ ($8|k\epsilon |\;K_{1}(|k\epsilon
|)=0.0496$), the first three existence intervals for the broad solitons are
\begin{equation}
2.17<\theta _{1}<2.62,\;\;\;5.30<\theta _{1}<5.73,\;\;\;8.43<\theta
_{1}<8.87,  \label{eq22}
\end{equation}
cf. Eq. (\ref{eq20}), with the strongest DDI attraction attained at $\theta
_{1}=$ $2.38,5.51$ and $8.65$, respectively.

\item[3)] In the case of Eq. (\ref{eq09}), when the unmodulated DDI
vanishes, the expansion converges slowly, so a numerical analysis of the
full integral expression in the GP equation (\ref{eq06})\ is necessary. In
this case, the solitons exist in much larger intervals of $\theta _{1}$ than
in the two previous cases. The first two intervals are

\begin{equation}
0.61<\theta _{1}<3.83,\;\;\;4.16<\theta _{1}<7.01.  \label{eq23}
\end{equation}
\end{itemize}

Further, the consideration of the three above cases reveals overlapping
regions where more than one soliton may exist:

\begin{eqnarray}
2.17 &<&\theta _{1}<2.29;\;\;\;2.52<\theta _{1}<2.62;\;\;\;5.30<\theta
_{1}<5.40;\;\;\;  \nonumber \\
5.63 &<&\theta _{1}<5.73;\;\;\;8.43<\theta _{1}<8.54.  \label{eq24}
\end{eqnarray}
Comparison of these predictions with results of the numerical solution of
the GP equation is given in Section IV.

\subsection{Variational approximation}

In order to apply the VA, we look for stationary solutions to Eq. (\ref{eq06}%
), with chemical potential $\mu $, as $\psi =\exp (-i\mu t)\phi (x)$, where $%
\phi (x)$ satisfies the equation
\begin{equation}
\mu \phi =-\frac{1}{2}\frac{d^{2}\phi (x)}{dx^{2}}+g|\phi (x)|^{2}\phi
(x)+G\phi (x)\int_{-\infty }^{+\infty }V_{\mathrm{DD}}(x,x^{\prime
})\left\vert \phi \left( x^{\prime }\right) \right\vert ^{2}dx^{\prime },
\label{eq25}
\end{equation}
which can be derived from the Lagrangian,
\begin{equation}
L=\int_{-\infty }^{+\infty }\mathcal{L}dx,  \label{eq26}
\end{equation}
where density $\mathcal{L}$ is given by
\begin{equation}
\mathcal{L}=\mu |\phi |^{2}-\frac{1}{2}\left\vert \frac{d\phi }{dx}
\right\vert ^{2}-\frac{g}{2}|\phi |^{4}-\frac{G}{2}|\phi
(x)|^{2}\int_{-\infty }^{+\infty }V_{\mathrm{DD}}(x,x^{\prime })|\phi
(x^{\prime })|^{2}dx^{\prime }.  \label{eq27}
\end{equation}

For a bright-soliton solution, we adopt the simplest Gaussian variational
ansatz, with amplitude $A$ and width $a$,
\begin{equation}
\phi =A\exp \left( -\frac{x^{2}}{2a^{2}}\right) ,  \label{eq28}
\end{equation}
whose norm (\ref{eq12}) is $N=\sqrt{\pi }A^{2}a$. The substitution of the
ansatz into Lagrangian (\ref{eq26}) with density (\ref{eq27}) yields the
corresponding averaged Lagrangian, which is expressed in terms of $N$,
instead of the amplitude $A$:
\begin{equation}
\bar{L}=N\left[ \mu -\frac{1}{4a^{2}}-\frac{gN}{2\sqrt{2\pi }a}-\frac{GN} {%
2\pi a^{2}}F\left( a,k,V_{0},V_{1}\right) \right] ,  \label{eq29}
\end{equation}
where we define
\begin{equation}
F\left( a,k,V_{0},V_{1}\right) \equiv \int_{-\infty }^{+\infty
}e^{-x^{2}/a^{2}}dx\int_{-\infty }^{+\infty }e^{-x^{\prime 2}/a^{2}} V_{%
\mathrm{DD}}(x,x^{\prime })dx^{\prime }  \label{eq30}
\end{equation}
[recall that $V_{\mathrm{DD}}(x,x^{\prime })$ is defined by Eqs. (\ref{eq05}
) and (\ref{eq07})]. Lastly, the Euler-Lagrange equations, $d\bar{L}/dN=$ $d
\bar{L}/da=0$, are derived from the averaged Lagrangian (\ref{eq29}):
\begin{eqnarray}
\mu &=&\frac{1}{4a^{2}}+\frac{gN}{\sqrt{2\pi }a}+\frac{GNF}{\pi a^{2}},
\nonumber \\
N &=&\frac{\sqrt{2\pi }}{-ga+\sqrt{2/\pi }G\left( a\partial F/\partial a-{2F}
\right) }.  \label{eq32}
\end{eqnarray}

\section{Numerical results}

In this section, we present results of the numerical solutions of the GP
equation (\ref{eq06}) for soliton modes, which are compared with predictions
based on the variational equations (\ref{eq32}).

\subsection{Dipoles oriented, on average, along the $x-$axis ($\protect%
\theta _{0}=0$)}

\begin{figure}[tbph]
\centerline{
\includegraphics[width=8cm,clip]{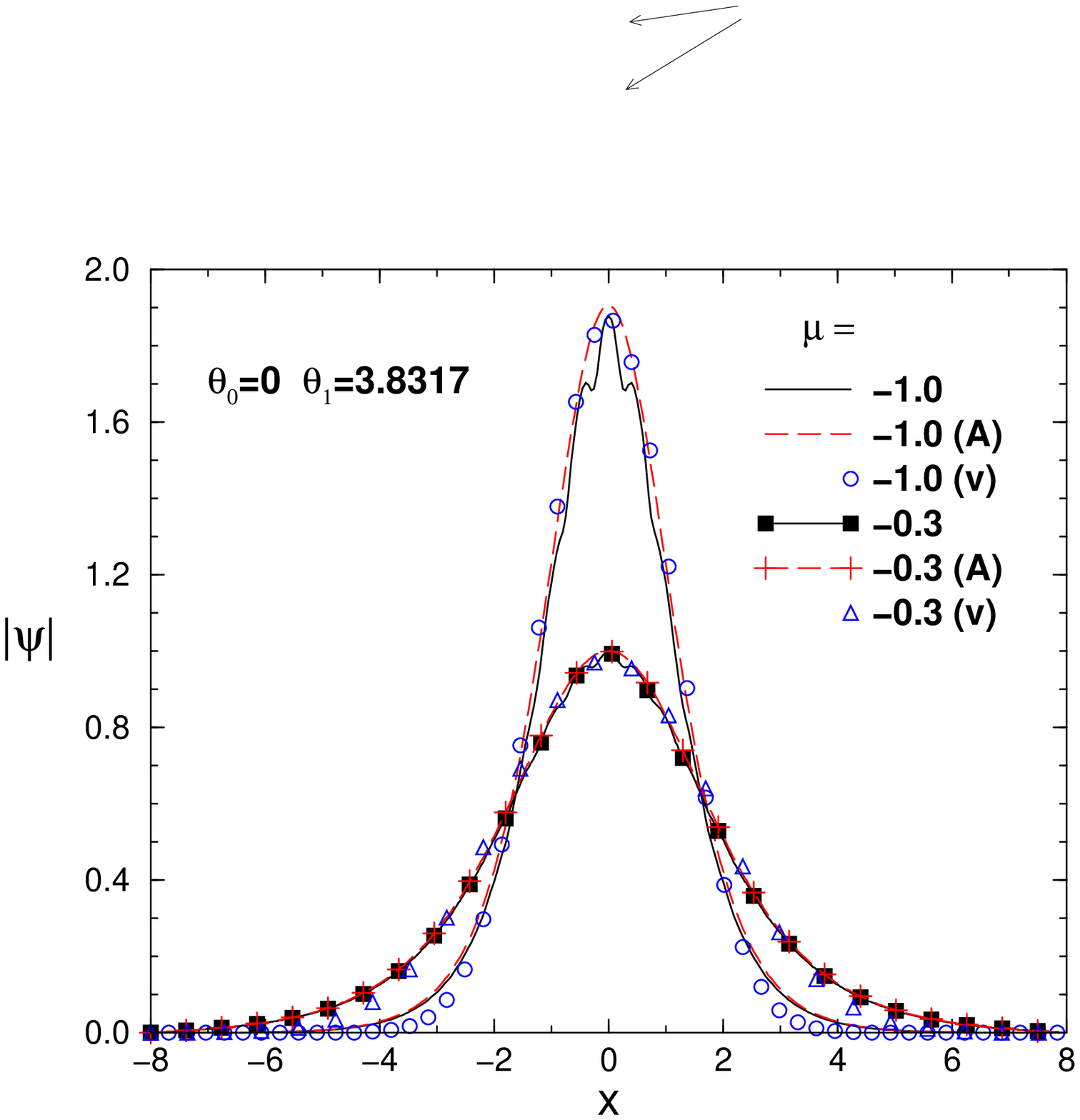}
}
\caption{(Color online) Soliton profiles for dipoles oriented, on the
average, parallel to the $x-$axis ($\protect\theta _{0}=0$), for $
\protect\theta _{1}=3.8317$, with $\protect\mu =-1.0$ and $-0.3$. Full
numerical solutions (solid lines, with solid squares pertaining to 
$\protect\mu =-0.3$) are compared with the approximate
solutions (A) obtained from Eq.~(\protect\ref{eq15}) (dashed lines, with
symbols plus indicating $\protect\mu =-0.3$) and with the
variational approximation (v).} 
\label{Fig-02}
\end{figure}
\begin{figure}[tbph]
\centerline{
\includegraphics[width=6cm,clip,angle=-90]{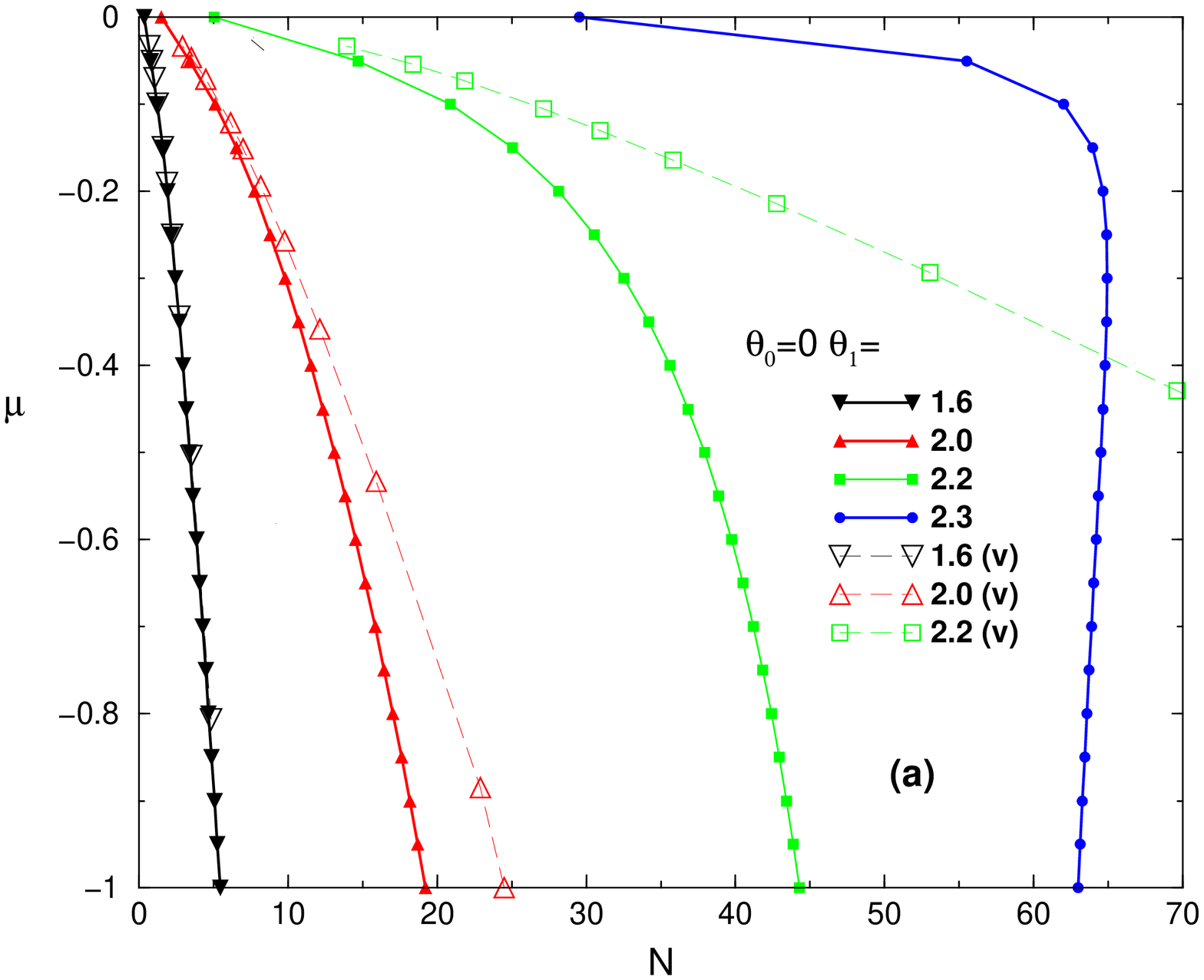}
\includegraphics[width=6cm,clip,angle=-90]{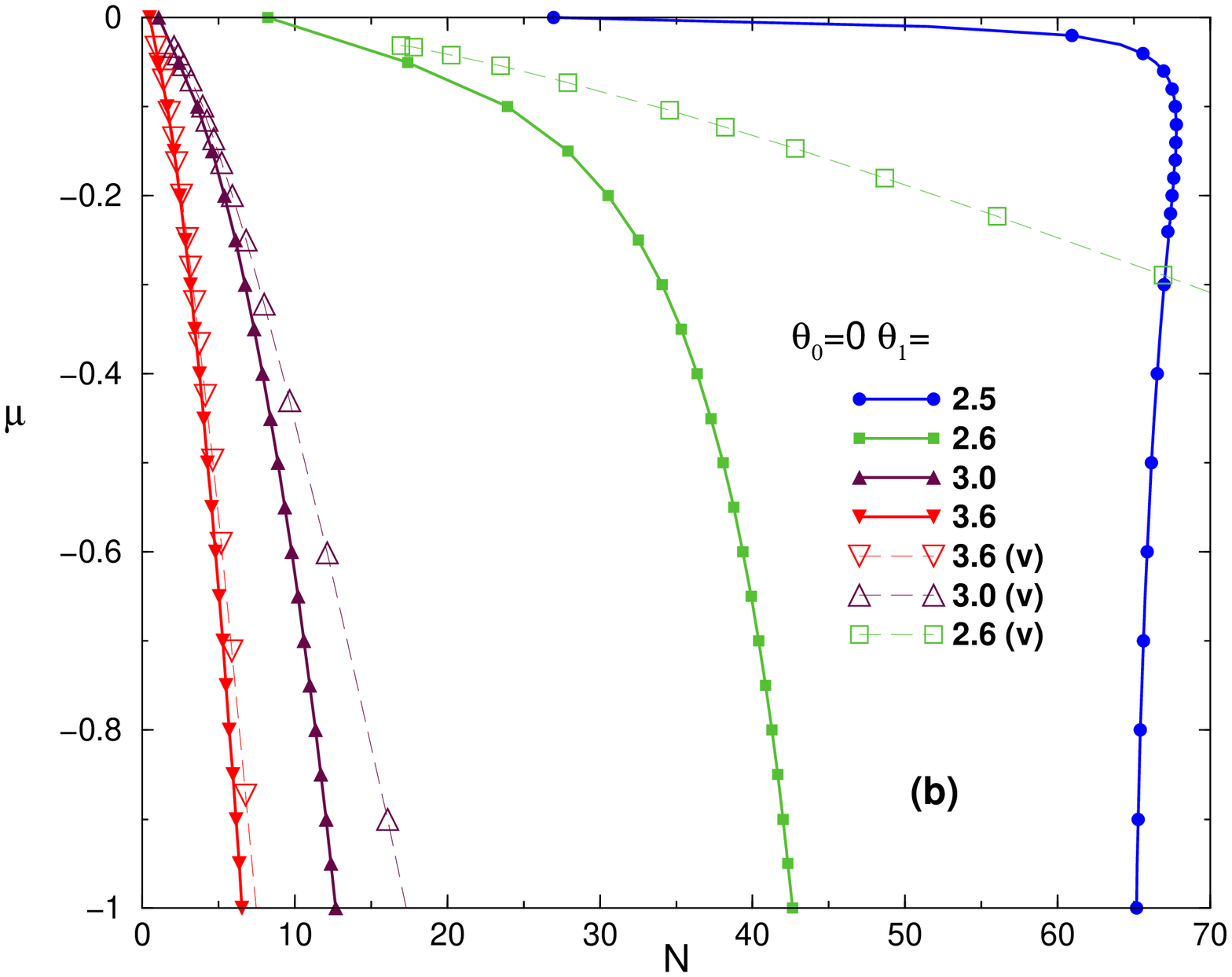}
}
\caption{(Color online) The chemical potential versus the soliton's norm,
for $\protect\theta _{0}=0$ with several values of $\protect\theta _{1}$,
from $1.6$ to $3.6$. The solutions are displayed in the intervals where the
stability is predicted by Eq.(\protect\ref{eq20}). The corresponding
variational results, indicated by (v), are also displayed, except in two
cases (close to unstable regions of parameters), where we cannot obtain
consistent results from the VA. }
\label{Fig-03}
\end{figure}
Numerically obtained solitary-mode density profiles for $\theta _{1}=3.8317$, 
which corresponds to a strong attractive DDI in the case of $\theta _{0}=0$, are
displayed in Fig.~\ref{Fig-02} for two values of the chemical potential: 
$\mu =-1$ and $-0.3$. The solutions were constructed using the full kernel 
(\ref{eq05}), the Bessel approximation (\ref{eq14}), as well as the
variational ansatz given by Eq.~(\ref{eq28}) with the amplitude and width
found from a numerical solution of algebraic equations (\ref{eq32}). A good
agreement is observed between the solutions of all the three types. 

In Fig.~\ref{Fig-03}, two panels are displayed for $\theta _{0}=0$,
with the chemical potential given as a function of the norm for the
orientation of the dipoles parallel to the $x\ $axis, for different values
of the the modulation amplitude $\theta _{1}$ in Eq.~(\ref{eq07}). The
Vakhitov-Kolokolov (VK) stability criterion, $d\mu /dN<0$~\cite{VK} (in
particular, the application of the VK criterion to nonlinear OLs was
developed in Ref.~\cite{AGST}) suggests that almost all the
localized modes are stable in the region of $0<\theta _{1}<3.6$,
except for a small region of parameter $\theta _{1}$. The unstable region,
given by $2.3<\theta _{1}<2.5$, can be extracted from the results presented
in panels (a) and (b). In panel (a), for
$\theta _{1}=2.3$, we observe a change of the slope indicating the loss of
stability for $\mu <-0.2$. Accordingly, in panel (b), this change of slope
can be observed for $\theta _{1}=2.5$, implying the loss of the stability at
$\mu <-0.1$. In both limiting cases, we have $N\approx 65$.
The VA is, in general, consistent in predicting the stability by means of
the VK criterion, even considering that the agreement of the predicted shape
of the solutions with their numerical counterparts becomes poor for large
values of $N$ and $\mu $. We also note that, close to the region where we
have unstable full numerical results (for large $\mu $) we are not able to
reach any conclusion by means of the VA, therefore only full numerical
results are presented for $\theta _{1}=$ $2.3$ and $2.6$. In general, for
the stable cases, the comparison with the predictions of the VA shows that
the agreement is rather good, with the discrepancy amounting to small shifts
in the values of $N$.

Following the results presented in Figs.~\ref{Fig-02} and 
\ref{Fig-03} for $\theta _{0}=0$, at larger values of this angle,$
0<\theta _{0}<\pi /4$, we have observed a similar stable behavior,
produced by numerical solutions and the variational analysis.
In general, it is observed that, by increasing the mean angle $\theta _{0}$
from zero, the existence region of stable solitons shrinks. In the next two
subsections we consider two other cases of interest.

\subsection{Dipoles oriented, on average, perpendicular to the $x-$axis 
($\protect\theta _{0}=\protect\pi /2$)}

In the case of $\theta _{0}=\pi /2$, when the dipoles are oriented, on
average, perpendicular to the $x\ $axis, stable solitons are found only in a
small interval of parameter $\theta _{1}$. Indeed, by using the VK
criterion, one can conclude that almost all values of $\theta _{1}$
correspond to unstable states, except for $\theta _{1}$ smaller than $%
\approx 1.5$.

\begin{figure}[tbph]
\centerline{
\includegraphics[width=8cm,clip]{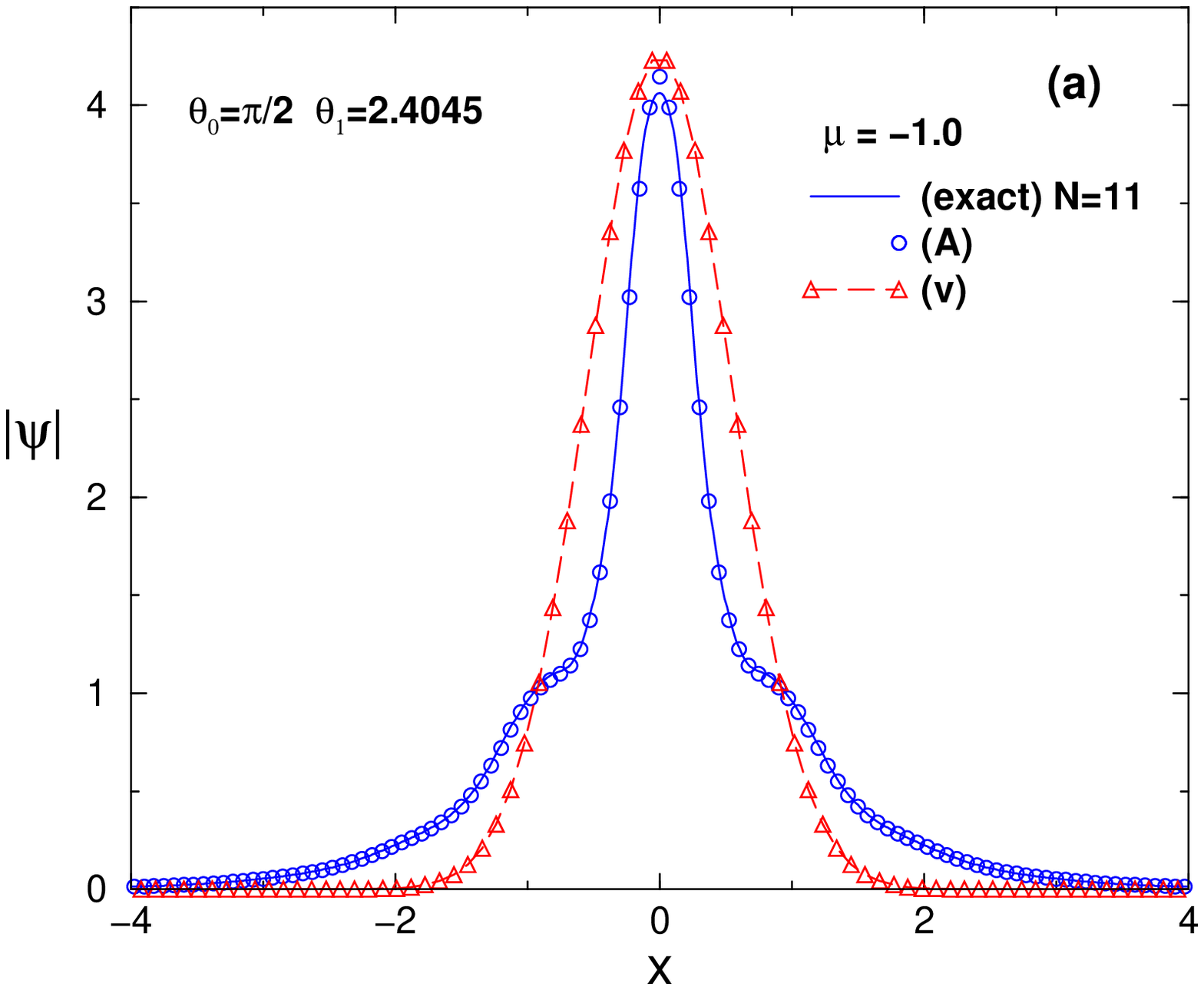}
\includegraphics[width=8cm,clip]{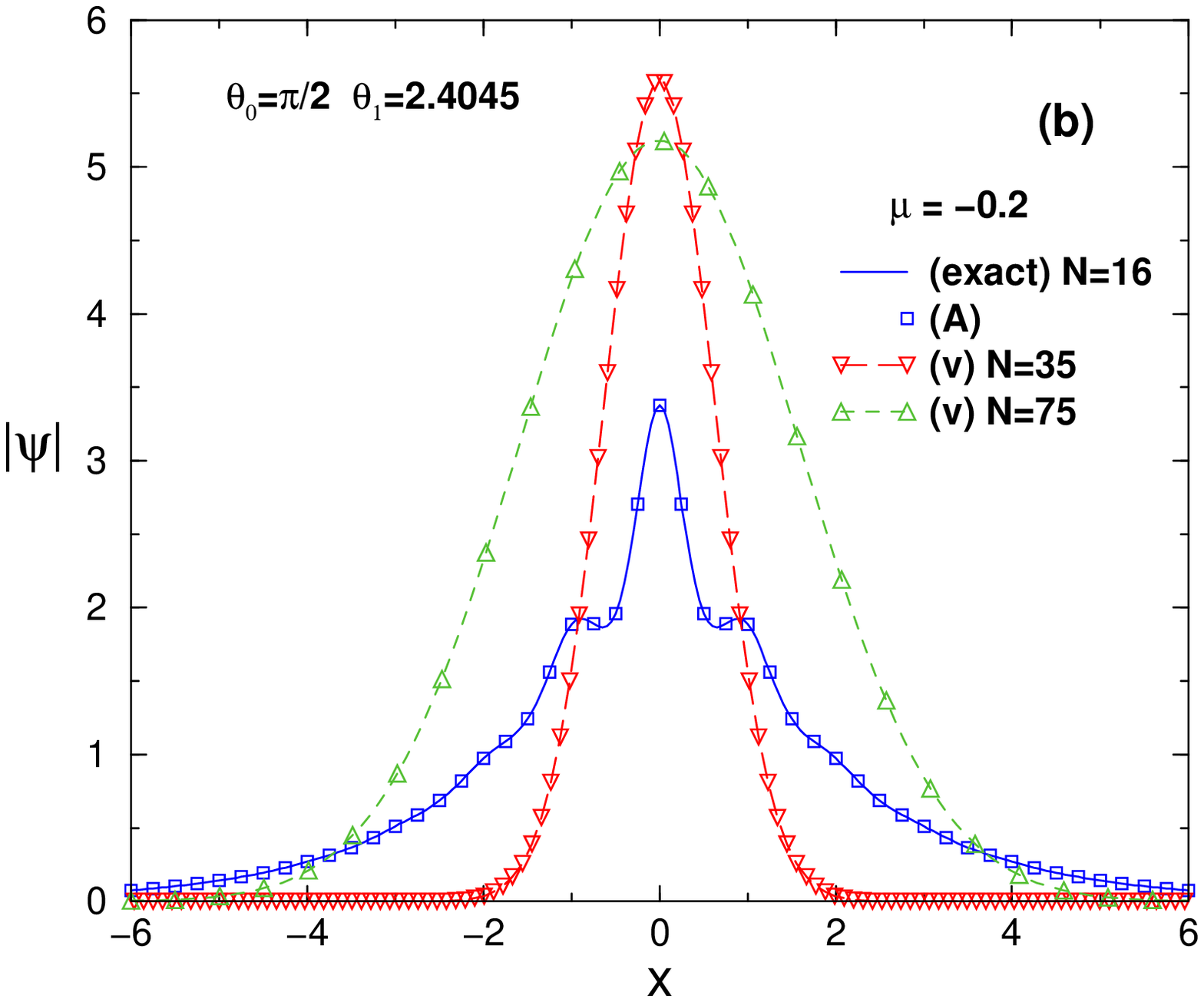}
}
\caption{(Color online) Soliton profiles for the dipoles oriented, on the
average, perpendicular to the $x-$axis ($\protect\theta _{0}=\protect\pi /2$
), with $\protect\mu =-1$ and $N\approx 11$ [panel (a)] and $\protect\mu %
=-0.2$ [panel (b)]. The full-numerical results are shown by solid (blue)
lines. The results with the approximate kernel (\protect\ref{eq16}), shown
by squares, are in perfect agreement with exact numerical results. The VA
results (v) are shown by dashed (red) lines and up-triangles. In (b), we
also show, by small dashed (green) lines and down-triangles, that another VA
solution exists, such that we have two possible values for the norm, $%
N\approx 35$ and $N\approx 75$ as indicated, both being far from the exact
one, which is $N\approx 16$.}
\label{Fig-04}
\end{figure}

In Fig.~\ref{Fig-04}, we display two panels for soliton profiles, with
chemical potentials $\mu =-1$ (a) and $\mu -0.2$ (b), for $\theta _{1}=2.408$%
, which corresponds to the strongest attractive DDI. A good agreement is
observed between the results produced by the approximate kernel (\ref{eq16})
(where the Bessel expansion is used) and the full numerical solutions. As
concerns the comparison of the numerically exact results with the VA
predictions, the Gaussian ansatz (\ref{eq28})\ may be too simple in this
case to reproduce the exact results. Actually, the Gaussian profile of the
VA follows the exact solution approximately in the case of $\mu =-1.0$,
shown in (a). But, for $\mu =-0.2$, two solutions with different values of $N
$ are predicted by the VA for the same $\mu $, both being very inaccurate,
as shown in the frame (b).

\begin{figure}[tbph]
\centerline{
\includegraphics[width=8cm,clip]{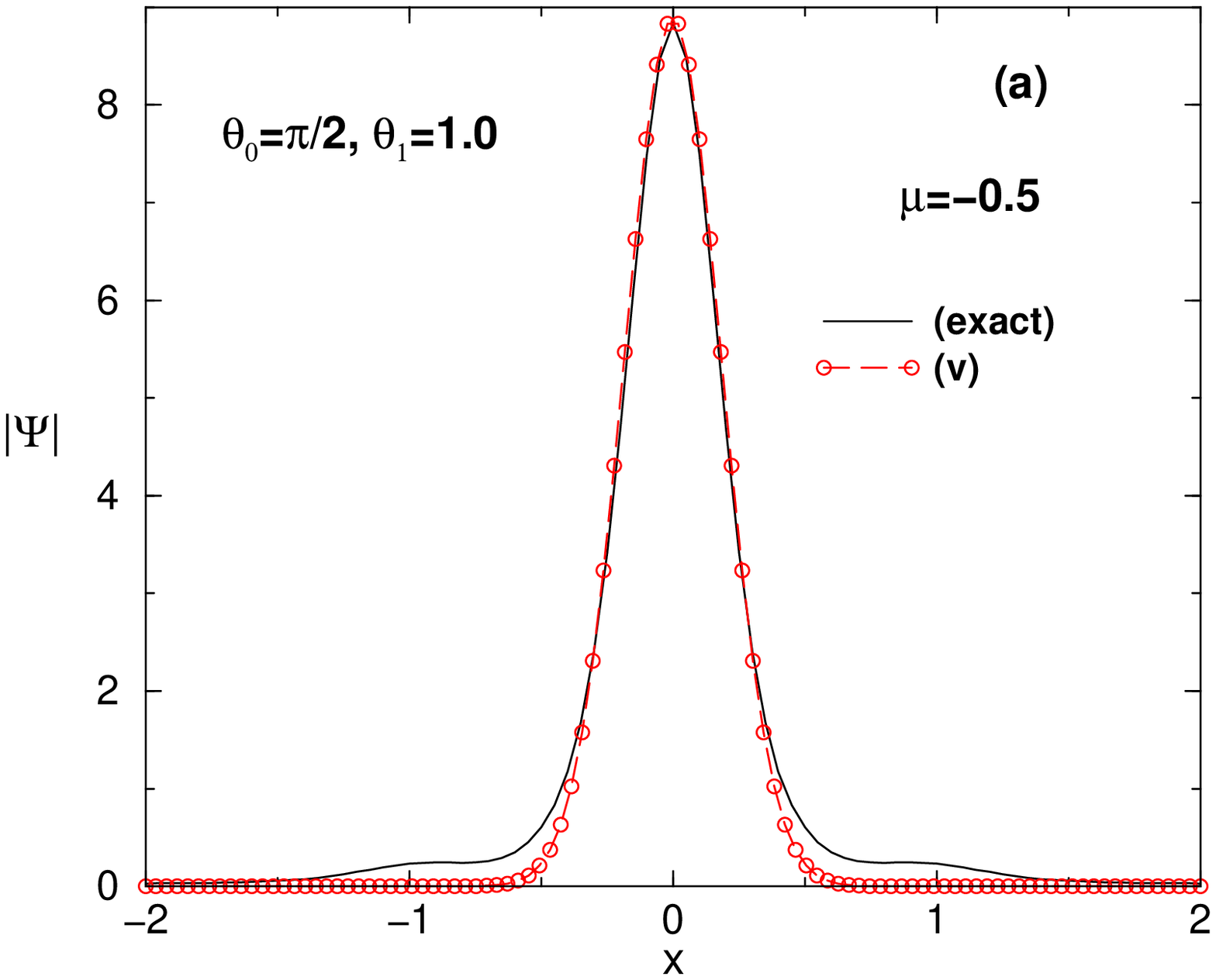}
\includegraphics[width=8cm,clip]{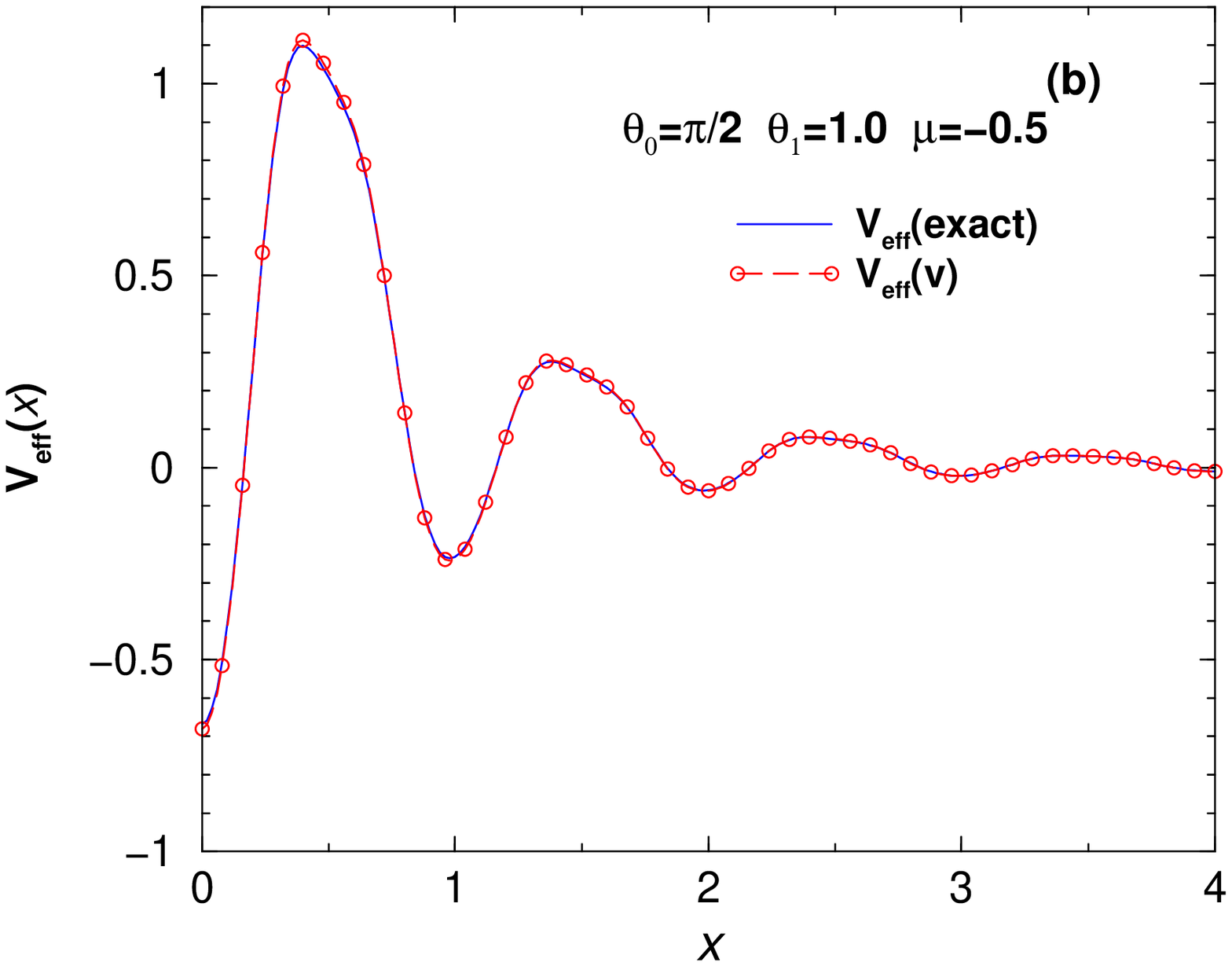}
}
\caption{(Color online) The soliton profile, for $\protect\theta _{0}=%
\protect\pi /2$, $\protect\theta _{1}=1.0$ and $\protect\mu =-0.5$, is shown
in the panel (a), with the corresponding effective potential, $V_{\mathrm{eff
}}(x)$, as defined by Eq.(\protect\ref{eq11}) with $g=0$ and $G=1$, shown in
the panel (b). Numerically exact results are given by solid (black) lines.
The corresponding variational (v) calculations are represented by dashed
(red) lines with circles. }
\label{Fig-05}
\end{figure}

The effective (pseudo) potential $V_{\mathrm{eff}}(x)$, which is defined by
Eq.(\ref{eq11}), plays an important role in the analysis of the soliton
modes. To display characteristic profiles of the potential, in 
Fig.~\ref{Fig-05} we plot the soliton profile [panel (a)] and the corresponding
effective potential [panel (b)] for $\theta _{1}=1.0$ and $\theta _{0}=\pi /2
$, with $\mu =-0.5$. In this case, the solutions are stable,
featuring a very close agreement between the  numerical and
variational results.
\begin{figure}[tbph]
\centerline{
\includegraphics[width=4.2cm,clip]{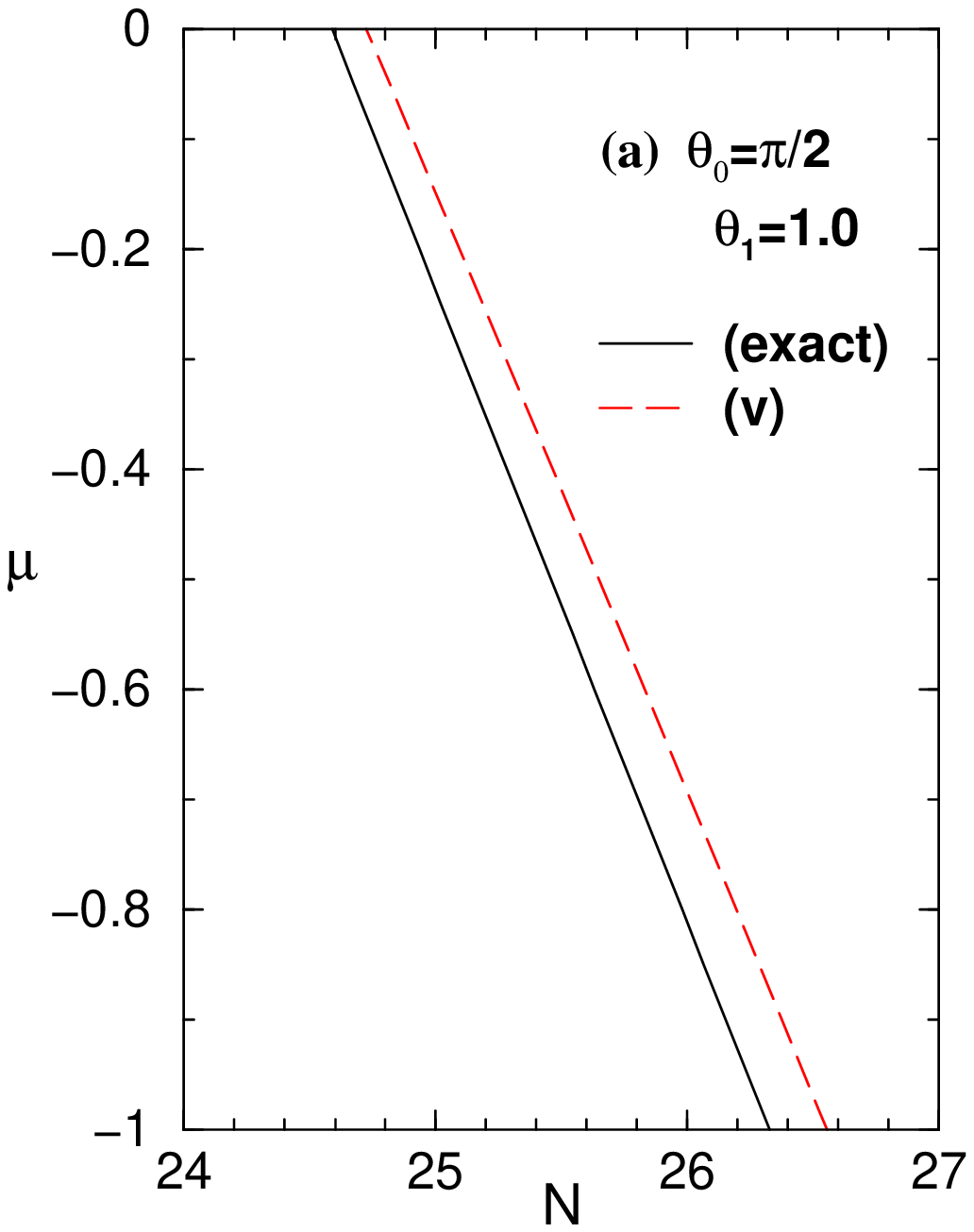}
\includegraphics[width=6cm,clip]{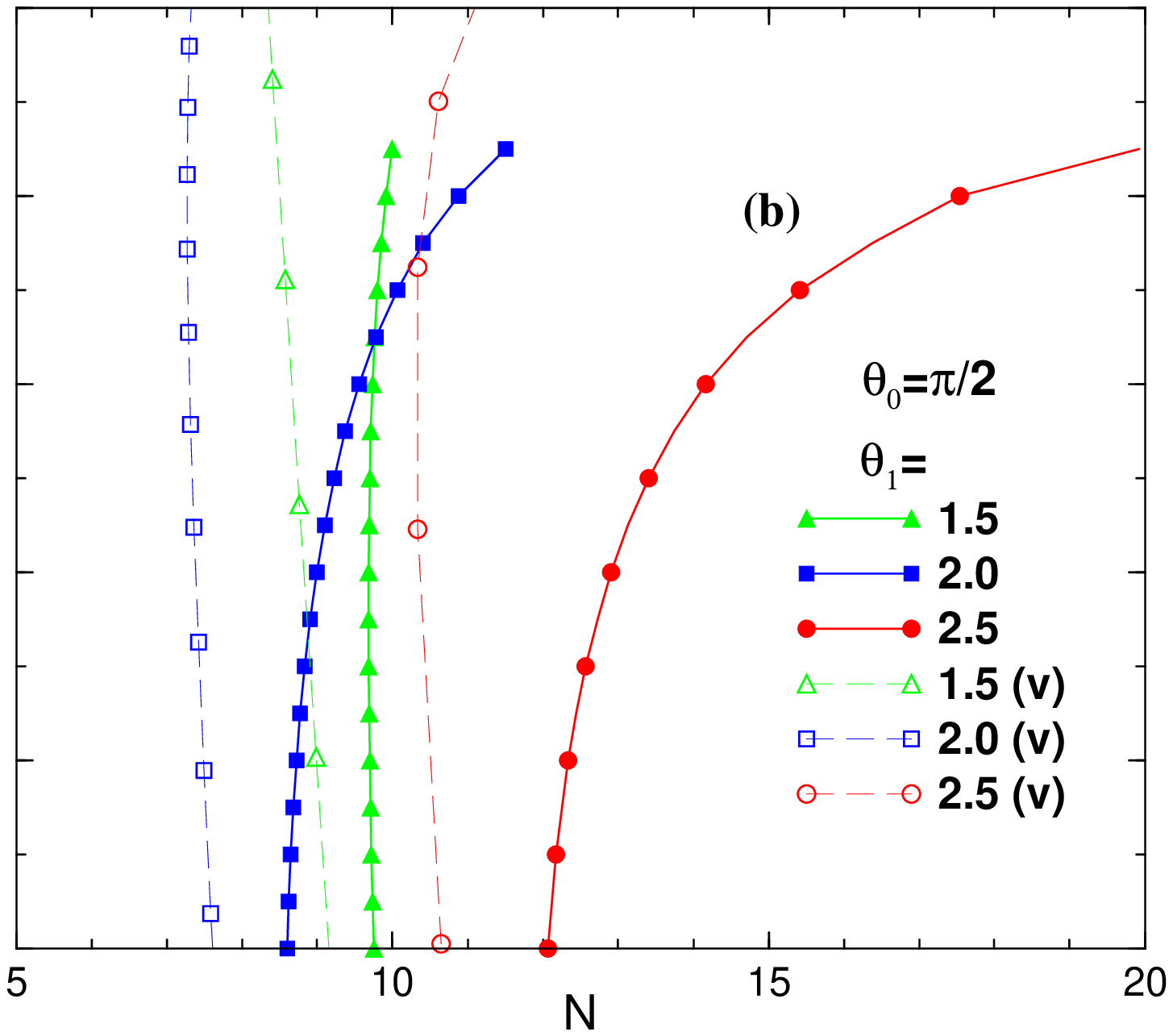}
\includegraphics[width=6cm,clip]{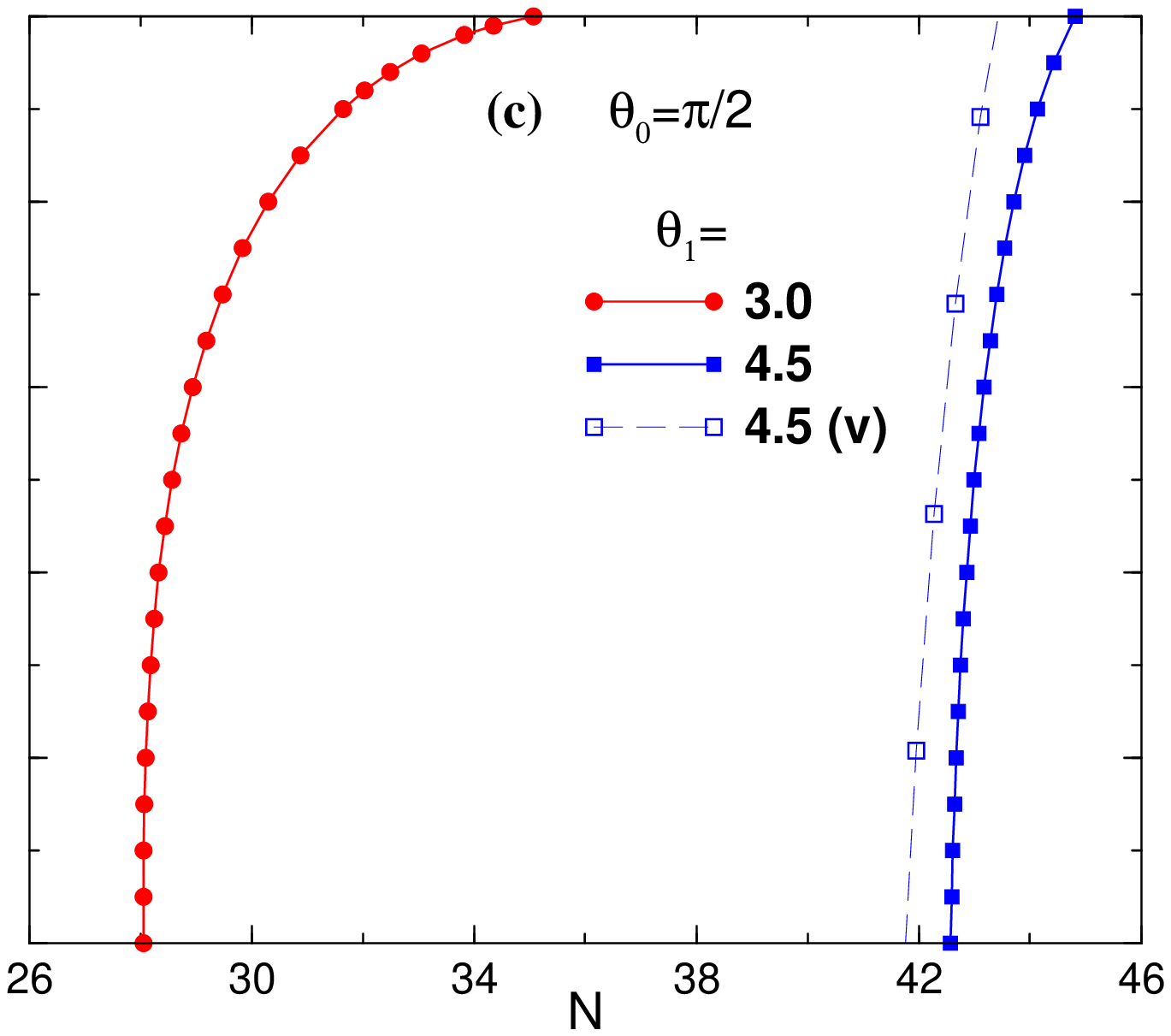}
}
\caption{(Color online) The chemical potential versus the norm, for $\protect
\theta _{0}=\protect\pi /2$ and several values of $\protect\theta _{1}$. The
variational results, indicated by label (v), provide a reasonable agreement
with the numerical results as $|\protect\mu |$ increases, and only
for values of $\protect\theta _{1}$ which do not belong to the interval of $
2.5<\protect\theta _{1}<4$. In panel (c), for $\protect\theta _{1}=3.0$,
there is no VA solution in the displayed interval of $\protect\mu $.}
\label{Fig-06}
\end{figure}

In Fig.~\ref{Fig-06}, we present three panels with numerical
results for the dependence between the chemical potential and norm at
several values of $\theta _{1}$. In terms of the VK criterion,
within the interval of $\mu $ presented here, $-1<\mu <0$, the stability of
the numerical solutions is quite clear only for $\theta _{1}=1.0$, as shown
in frame (a). As we increase $\theta _{1}$, in panel (b) we see that
the system is already at the stability limit for $\theta _{1}=1.5$, and
becomes unstable for larger values of this parameter. The stability
predictions of the VA, which are also displayed in the figure, are in
reasonable agreement with the numerical results
for larger values of $|\mu |$, where both the VA and full numerical
solutions produce similar slopes. This conclusion is also supported
by the perfect agreement between the variational and numerical
results in Fig.~\ref{Fig-05}. For other solutions, presented in the panels
(b) and (c) of Fig.~\ref{Fig-06}, the VK criterion does not predict 
clear stability regions. The agreement between the VA and full numerical
solutions is completely lost in the interval of $2.5<\theta _{1}<4$, where
no reasonable VA solutions was found for $-1<\mu <0$. For the sake of the
comparison with the VA, we keep the plot with $\theta _{1}=2.5$ in panel
(b), which shows how the agreement between the VA and numerical data starts
to deteriorate. Therefore, for $\theta _{1}=3.0$, we can only show the
exact numerical results. Another point to be observed in these plots is that
a conspicuous region of the reasonable agreement of the VA with numerical
results (although without stabilization of the solitons) reappears at large
values of $\theta _{1}$, as seen in the panel (c) for $\theta _{1}=4.5$.

\begin{figure}[tbph]
\centerline{
\includegraphics[width=8cm,clip]{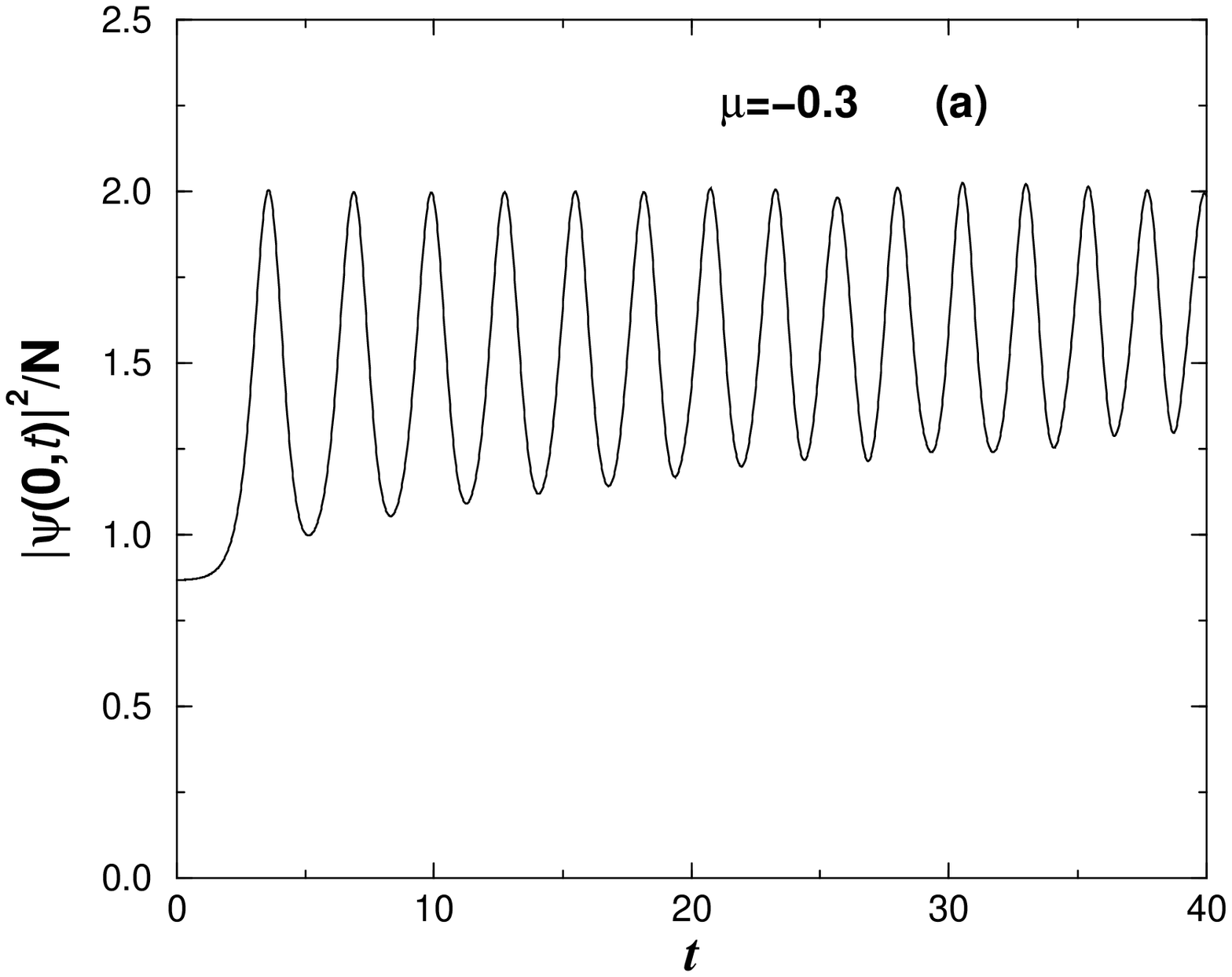}
\includegraphics[width=7.5cm,clip]{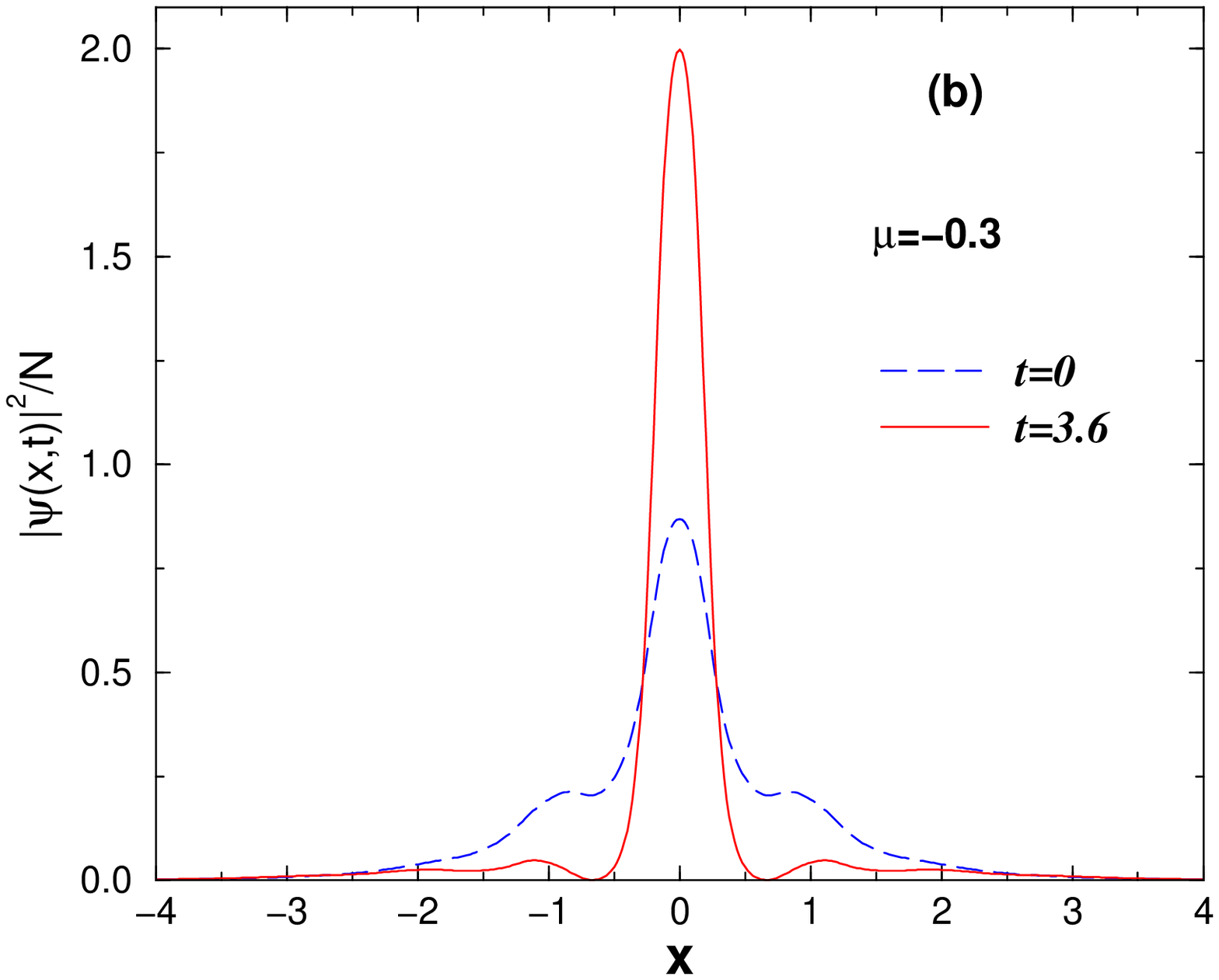}
}
\caption{(Color online) A typical example of the evolution of the peak
density [panel (a)] in the case when unstable solitons do not decay,
featuring instead oscillations between two different shapes, as shown in the
panel (b). Here, $\protect\theta _{0}=\protect\pi /2$ and 
$\protect\theta_{1}=2.4048$.}
\label{Fig-07}
\end{figure}

Concerning the evolution of unstable solitons, there are generic situations
in which the solitons are not exactly stable, but they do not decay either.
This happens, for example, at $\theta _{0}=\pi /2$ and $\theta _{1}=2.4048$,
as shown in Fig.~\ref{Fig-07}, where the profiles are presented in
frame (b) for two instants of time, with $\mu =-0.3$. In frame (a) we show
the peak of the density at the origin ($x=0$). In this case, the unstable
soliton is transformed into a persistent breather oscillating between two
different configurations. While such breathers are characteristic modes of
the present system with the spatially modulated DDI, they have not been
observed in models of the dipolar BEC with the constant DDI.

\begin{figure}[tbp]
\centerline{
\includegraphics[width=12cm,clip]{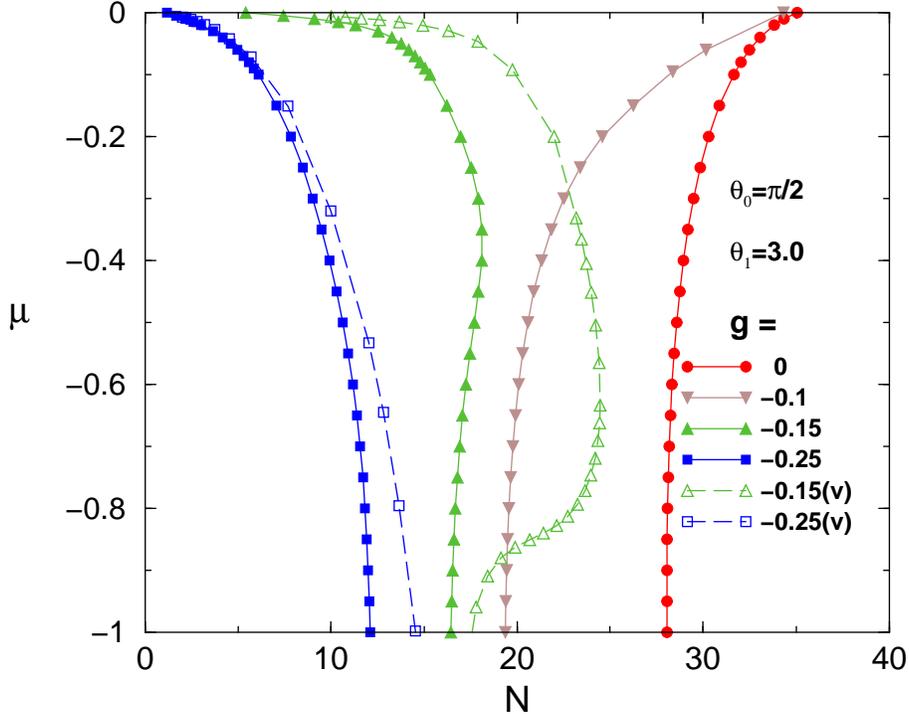}
}
\caption{(Color online) The role of the local self-attractive nonlinearity,
accounted for by coefficient $g<0$, in stabilizing the solitons at $\protect%
\theta _{0}=\protect\pi /2$ and $\protect\theta _{1}=3.0$ (which are shown
in Fig.~\protect\ref{Fig-06} to be unstable at $g=0$). As suggested by the
VK criterion, and is corroborated by numerical simulations (not shown here
in detail), the solutions are stabilized at $g<-0.15$. The corresponding
VA-predicted $\protect\mu (N)$ branches are shown for $g=-0.15$ and $-0.25$.
For smaller $|g|$, the simple Gaussian ansatz (\protect\ref{eq28})\ cannot
provide reliable results.}
\label{Fig-08}
\end{figure}

Next, we proceed to the consideration of the GP equation~(\ref{eq06}) which
includes the contact interactions, accounted for by $g\neq 0$. The results
are presented in Fig.~\ref{Fig-08} in terms of the $\mu (N)$ curves, which
show, by means of the VK criterion (whose validity has been tested in this
case by means of direct simulations), the role of this contact interaction
in stabilizing solitons that were found to be unstable for $g=0$, with $%
\theta _{0}=\pi /2$. To this end, we here consider the case of $\theta
_{1}=3.0$. The corresponding results for $g=0$ are presented above in panel
(c) of Fig.~\ref{Fig-06}, and are included here too for the sake of the
comparison. It is observed that the solutions get stabilized by the
self-attractive contact nonlinearity with $g<-0.15$. The results are
reported here for full numerical solutions, except in the cases where stable
solutions were found, where we also consider the corresponding VA (see the
cases of $g=-0.15$ and $-0.25$). At smaller $|g|$, the VA predictions are
far from the numerical results, at least for large values of $\mu $. A large
discrepancy between the variational and numerical curves behavior is
actually observed in Fig. \ref{Fig-08} for $g=-0.15$.

\subsection{The average dipoles' orientation corresponding to $\protect%
\theta _{0}=\protect\theta _{m}\equiv \arccos (1/\protect\sqrt{3})$}

\begin{figure}[tbph]
\centerline{
\includegraphics[width=10cm,clip]{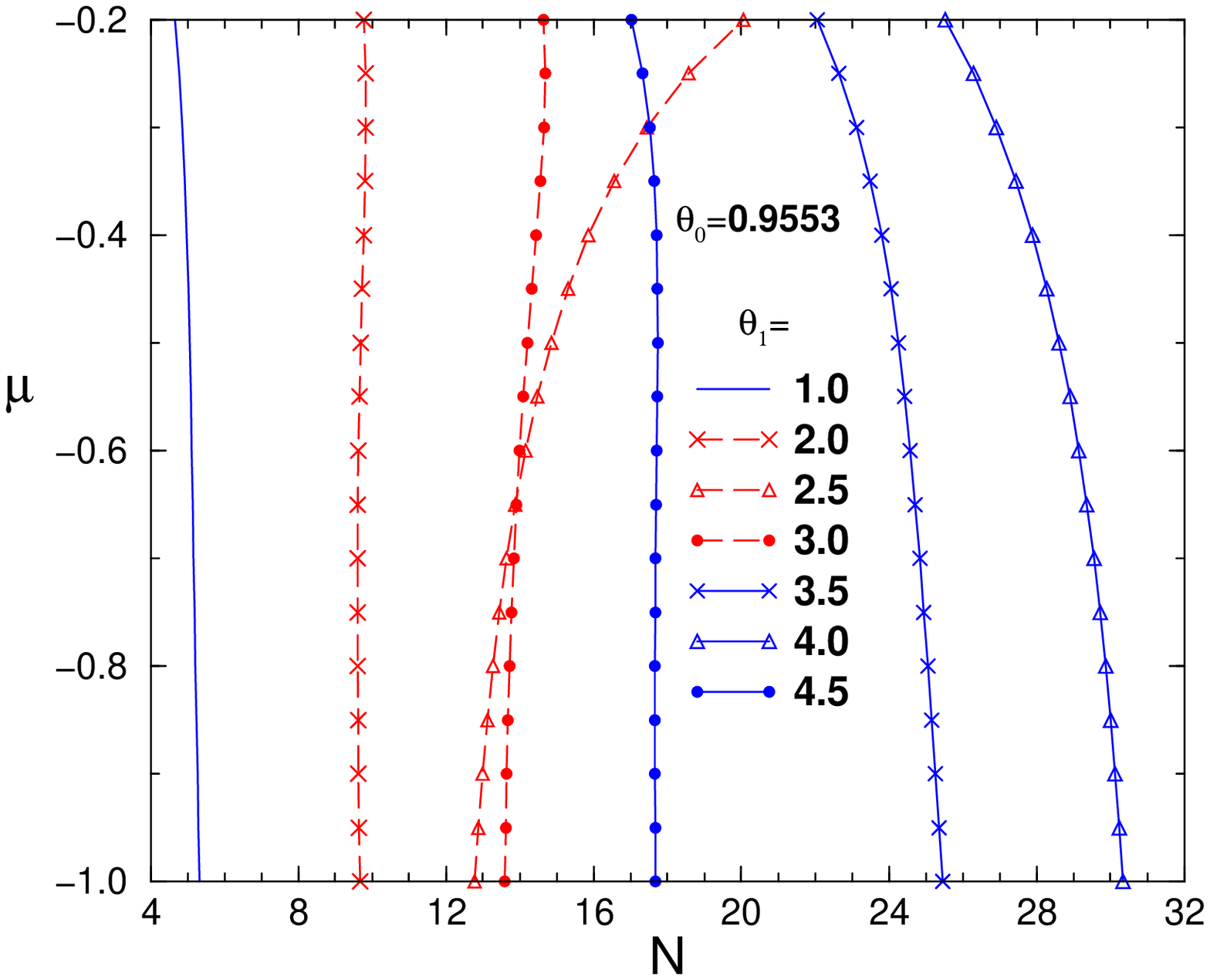}
}
\caption{(Color online) The chemical potential versus the norm, for 
$\protect\theta _{0}=\protect\theta _{m}\ $[see Eq. (\protect\ref{eq09})], 
at several
values of $\protect\theta _{1}$. In agreement with analytical predictions
for the region  of the soliton existence, see Eq.~(\protect
\ref{eq23}), in most cases we have stable results, except for the
small region of $2<\protect\theta _{1}<3$, where the results were
represented by dashed (red) lines. The respective variational results are
chiefly consistent, in terms of the stability prediction,
with the above, but not shown, as they yield much larger
values of $N$.}
\label{Fig-09}
\end{figure}

It is also interesting to consider the configuration with the mean
orientation of the dipoles corresponding to the zero interaction (if $\theta
_{1}=0$), as per Eq.~(\ref{eq09}). The corresponding dependence of $\mu $ on
$N$ is plotted in Fig.~\ref{Fig-09}, for several values of $\theta _{1}$,
from $1.0$ to $4.5$. In agreement with the predictions of the VA, in this
case most soliton solutions are stable, considering condition 
(\ref{eq23}). However,  the VA results yield much larger values of $N$, 
and it also predicts stable solitons in the region where the
numerical solutions are unstable, e.g., for $\theta _{1}=2.5$.

Summarizing our numerical analysis, exemplified by the above subsections for 
$\theta _{0}=0$, $\theta _{0}=\pi /2$ and $\theta _{0}=0.9553$, we conclude 
that the case of $\theta _{0}=0$ is the one where the stability is more likely 
to be reached for a wide range of values of the modulation parameter $\theta _{1}$, 
(see Fig.~\ref{Fig-03}). On the other hand, the case of $\theta _{0}=\pi /2$ is 
the one where the stability may occur only in a very specific region of values 
of $\theta _{1}$. 

\subsection{Soliton motion and collisions}

\begin{figure}[tbph]
\centerline{
\includegraphics[width=8cm,clip]{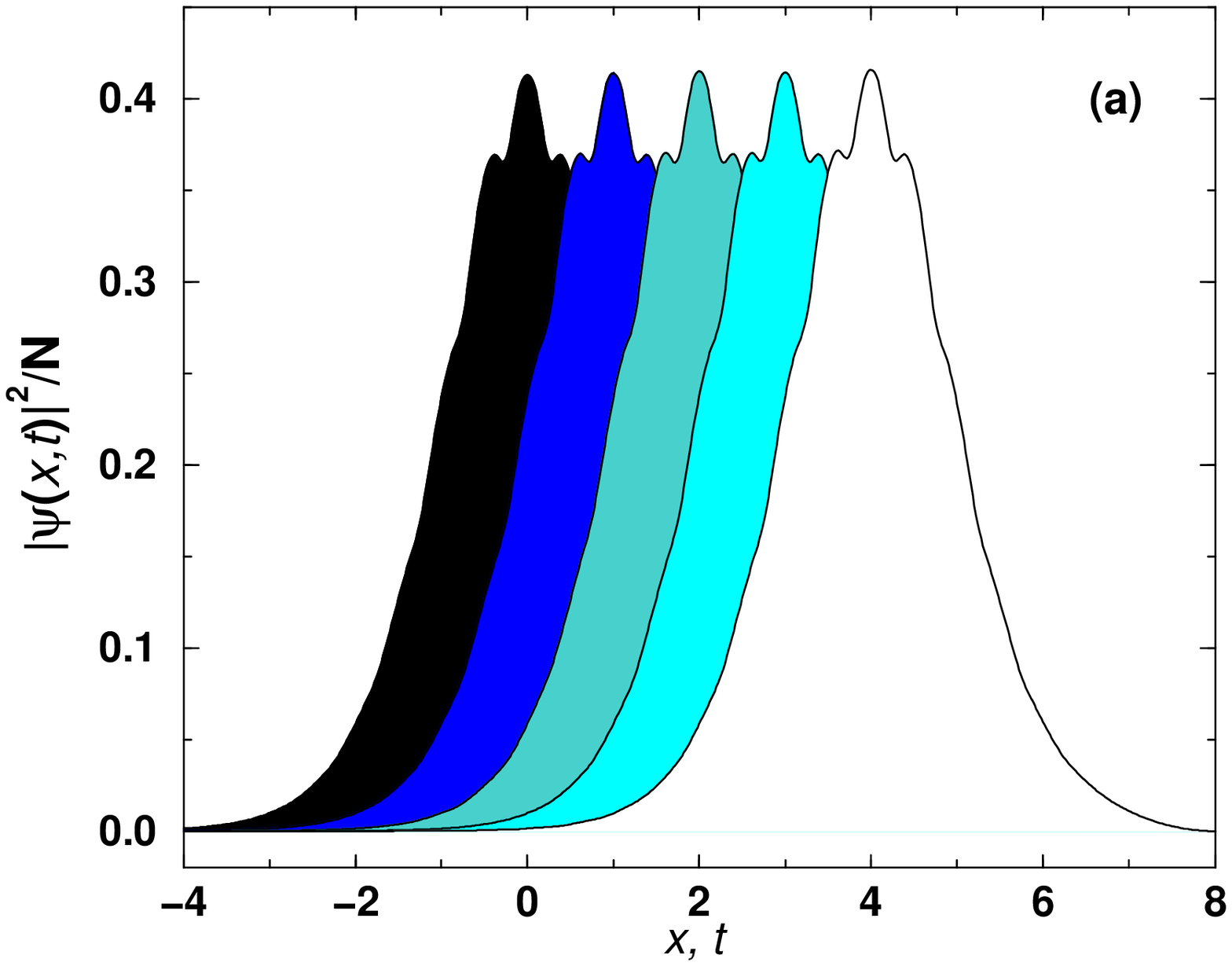}
\includegraphics[width=8cm,clip]{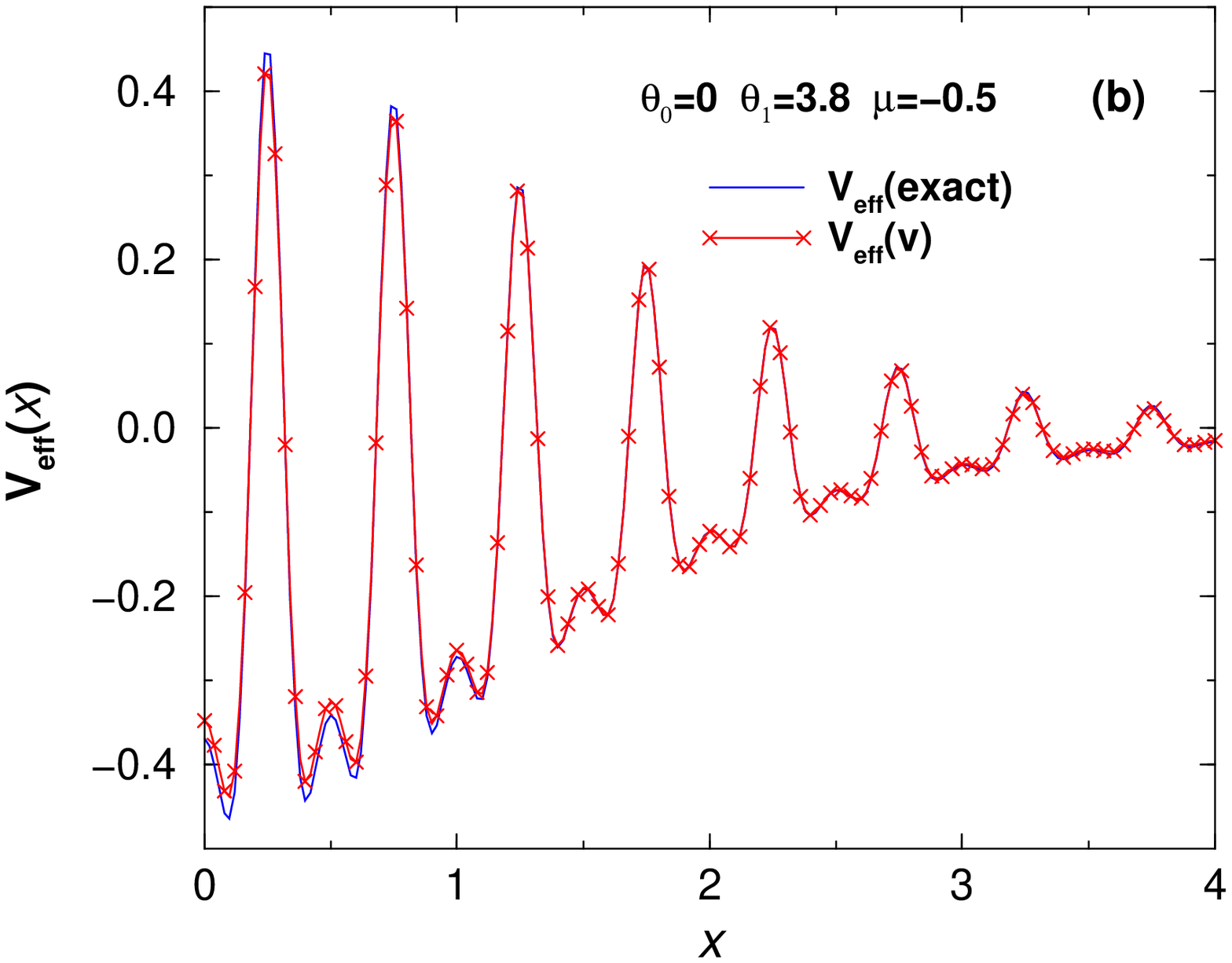}
}
\caption{(Color online) Numerical results for a soliton moving at velocity $%
v_{s}=1$ are shown in the panel (a), by means of the juxtaposition of
density profiles, with the horizontal axis simultaneously showing values of $%
x$ and $t$. The parameters are $\protect\theta _{0}=0$, $\protect\theta %
_{1}=3.8$ and $\protect\mu =-0.5$. In the panel (b), we have the
corresponding effective (pseudo) potential.}
\label{Fig-10}
\end{figure}

By means of numerical simulations, we have also investigated mobility of
solitons in the framework of the present model. As usual, the soliton was
set in motion, multiplying its wave function by the kicking factor, $\exp
\left( iv_{s}x\right) $. For a configuration with $\theta _{0}=0$, $\theta
_{1}=3.8$ and $\mu =-0.5$, and by taking $v_{s}=1$, our simulations
demonstrate the soliton propagation with a practically undistorted shape, as
shown in the panel (a) of Fig.~\ref{Fig-10}. Radiation losses are
practically absent in this case, as well as no excitation of internal modes
is observed in the soliton. In the panel (b) of Fig.~\ref{Fig-10}, the
effective potential, as given by Eq.~(\ref{eq11}), is plotted for the same
parameters. It gives rise to an effective Peierls-Nabarro potential
experienced by the moving soliton~\cite{KLakomy,AGT}. For a broad soliton,
whose width is large in comparison with the internal scale of the potential
(this is the case in Fig.~\ref{Fig-10}), the effective obstacle created by
the potential is exponentially small, hence one may expect practically free
motion of the soliton, which is confirmed by the simulations, even for an
initial speed reduced to $v_{s}=0.05$. At such a low speed too, the soliton
propagates along the lattice without changing its shape.

\begin{figure}[tbph]
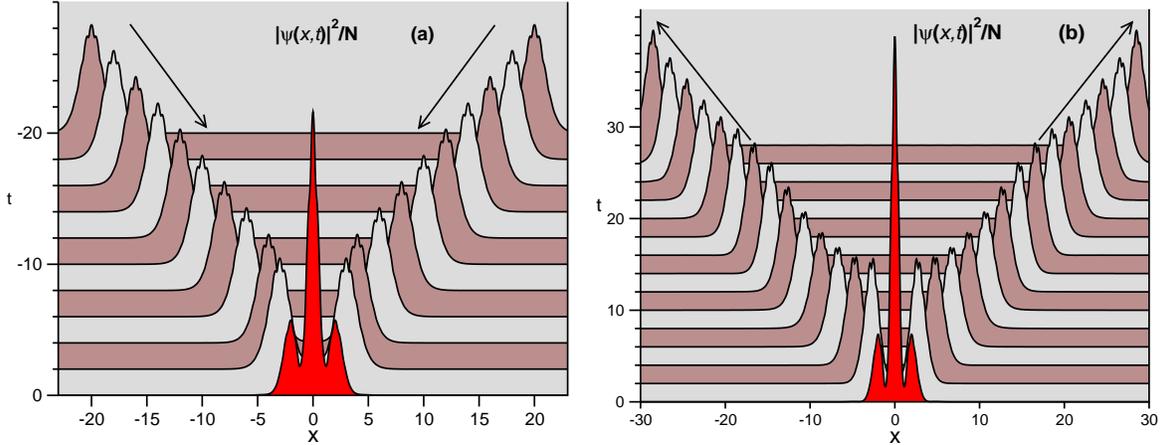

\centerline{
\includegraphics[width=7.5cm,clip]{fig11a.eps}
\hskip 0.2cm
\includegraphics[width=7.5cm,clip]{fig11b.eps}
}
\caption{(Color on-line) The two-soliton collision at the same parameters as
in Fig.\protect\ref{Fig-10}. Originally, the solitons' centers are placed at
points $x=\pm 20$, as shown in the panel (a), with the collision point being
$x=0$. The collision gives rise to internal excitations of the solitons,
which manifest themselves as periodic oscillations of their amplitudes,
observed in the panel (b). }
\label{Fig-11}
\end{figure}

Concluding the numerical analysis of the moving solitons, in 
Fig.~\ref{Fig-11} the collision of two solitons is displayed for the same set of
parameters as given in Fig. \ref{Fig-10}. Panels (a) and (b) display the evolution
of the density profiles, before and after the collision, respectively.
As one observes in panel (b), the collision does not destroy the solitons,
but rather induces fluctuations in the shapes of the colliding solitons
(probably related to excitation of internal modes), eventually showing an
almost elastic collision.

\begin{figure}[tbph]
\centerline{
\includegraphics[width=16cm,clip]{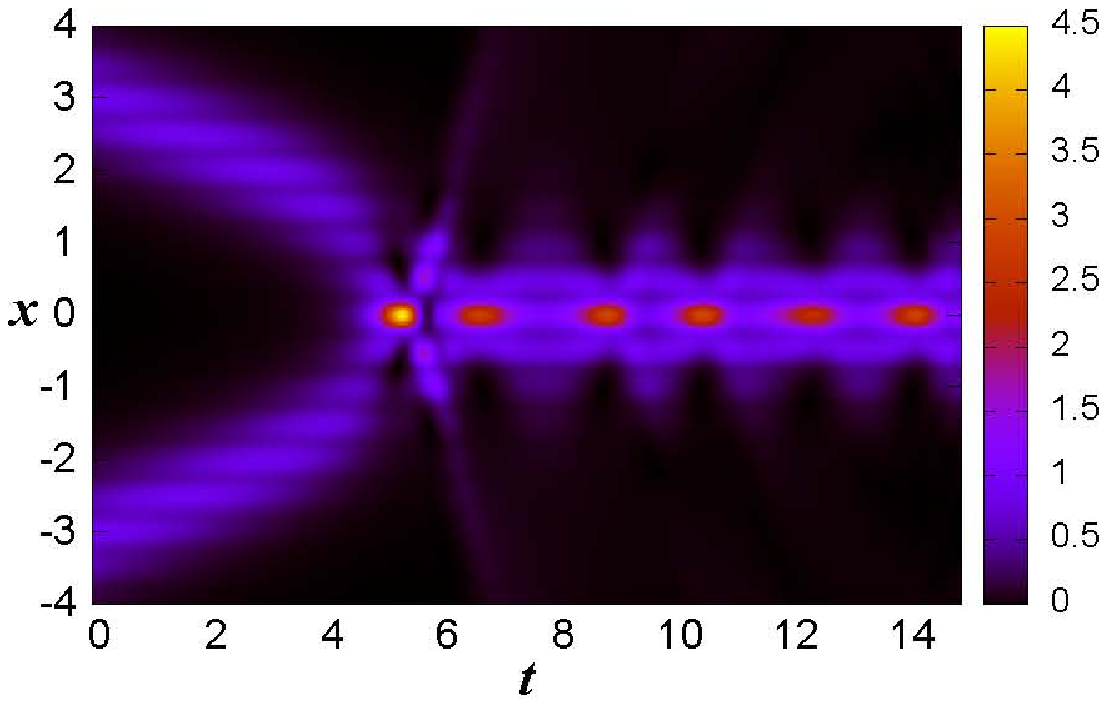}
}
\centerline{
\includegraphics[width=8.2cm,clip]{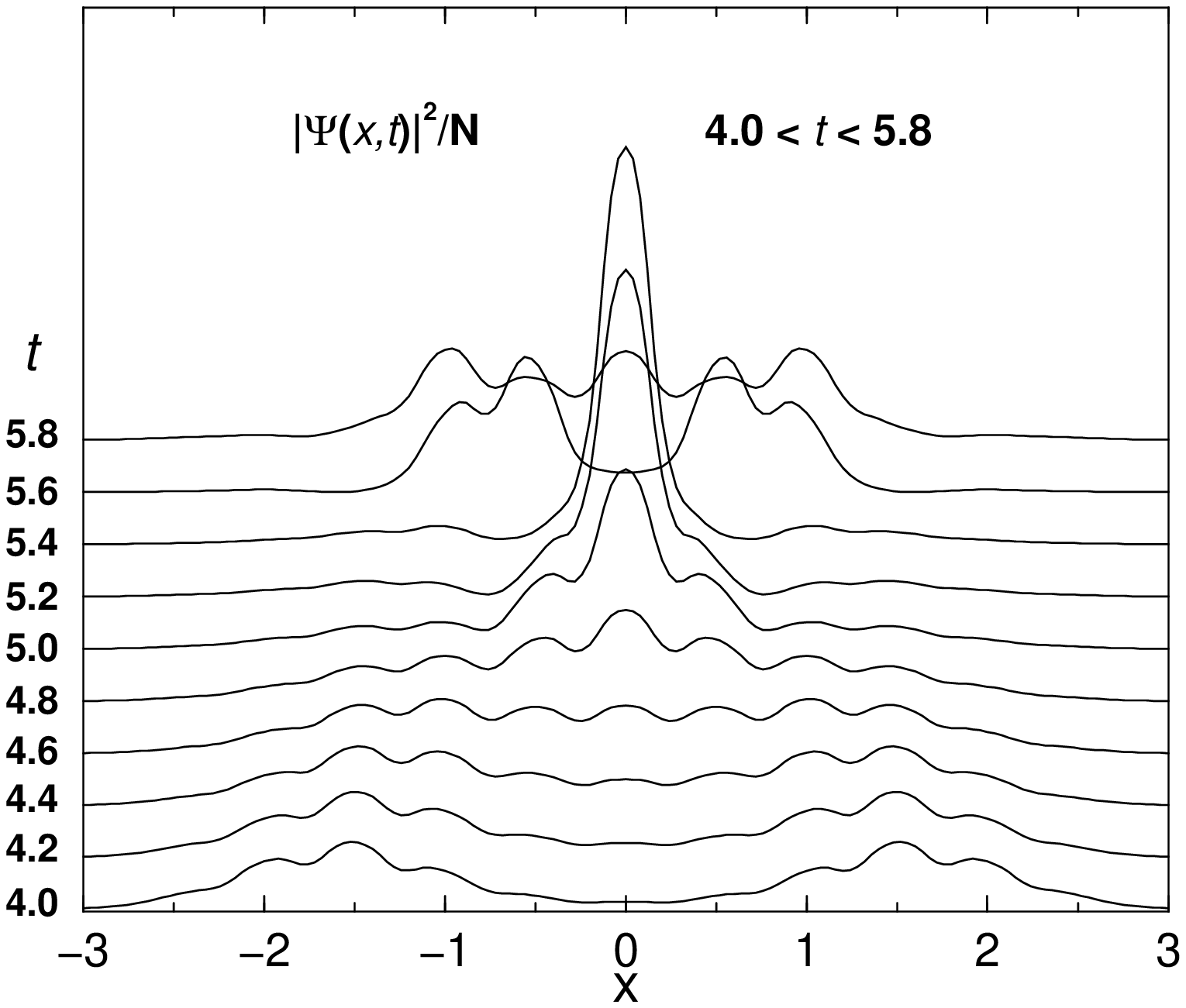}
\includegraphics[width=8cm,clip]{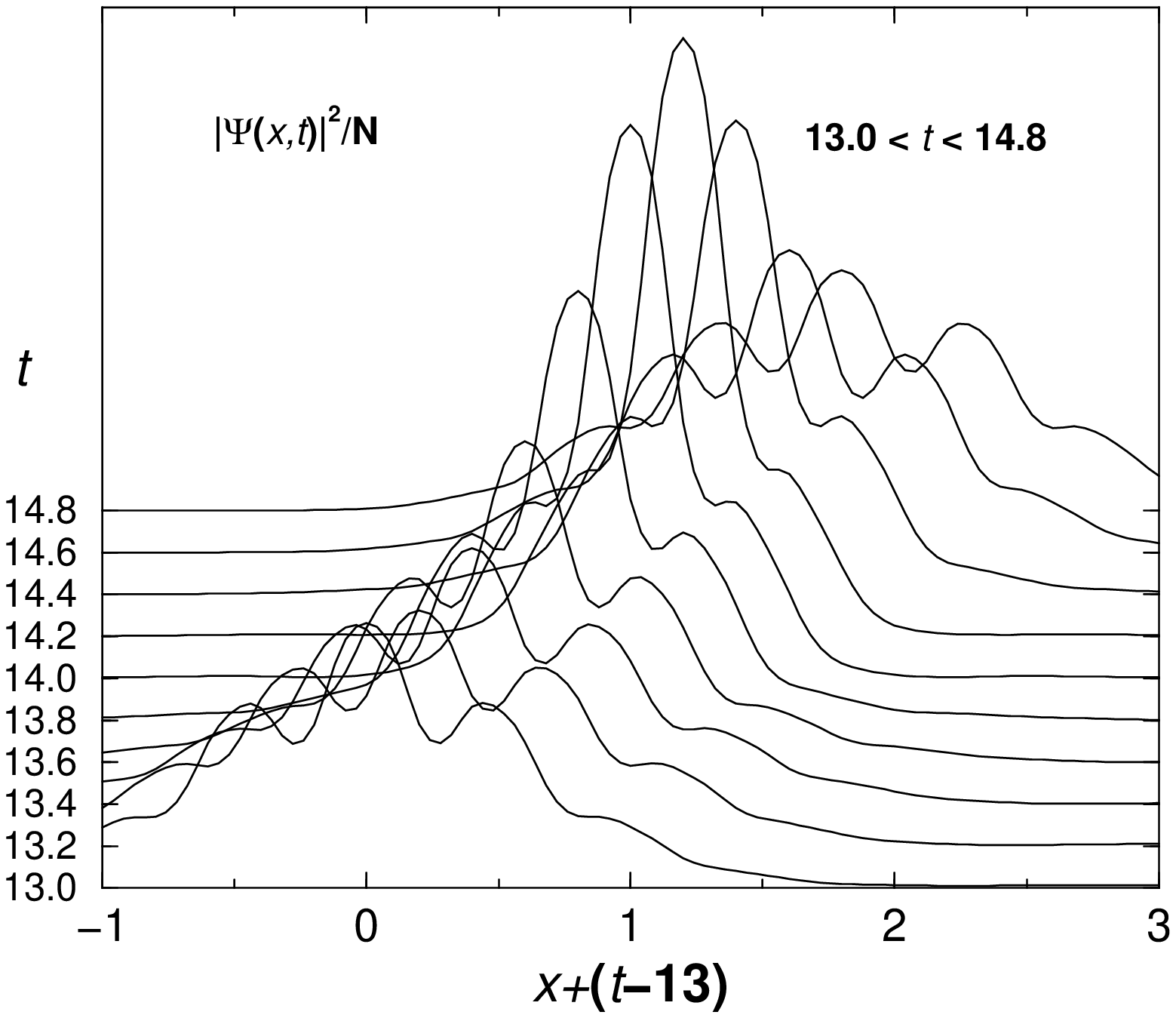}
}
\caption{ Illustrated in the upper panel, we present the evolution of
density profiles $|\protect\psi (x,t)|^{2}/N$, for the collision of two
solitons, for $\protect\theta _{0}=0$, $\protect\theta _{1}=3.8$, with
chemical potential $\protect\mu =-2.5$. The initial velocities and positions
of the solitons are $v_s \pm 0.3$ and $x=\mp 3$, respectively. The
juxtapositions of density profiles for their time evolution are shown in the
bottom rows, considering two time intervals, as shown inside the frames. The
formation of a bound state trapped around $x=0$, along with strong emission
of radiation, is clearly observed as the result of this inelastic collision.
}
\label{Fig-12}
\end{figure}

\begin{figure}[tbph]
\centerline{
\includegraphics[width=7.5cm,clip]{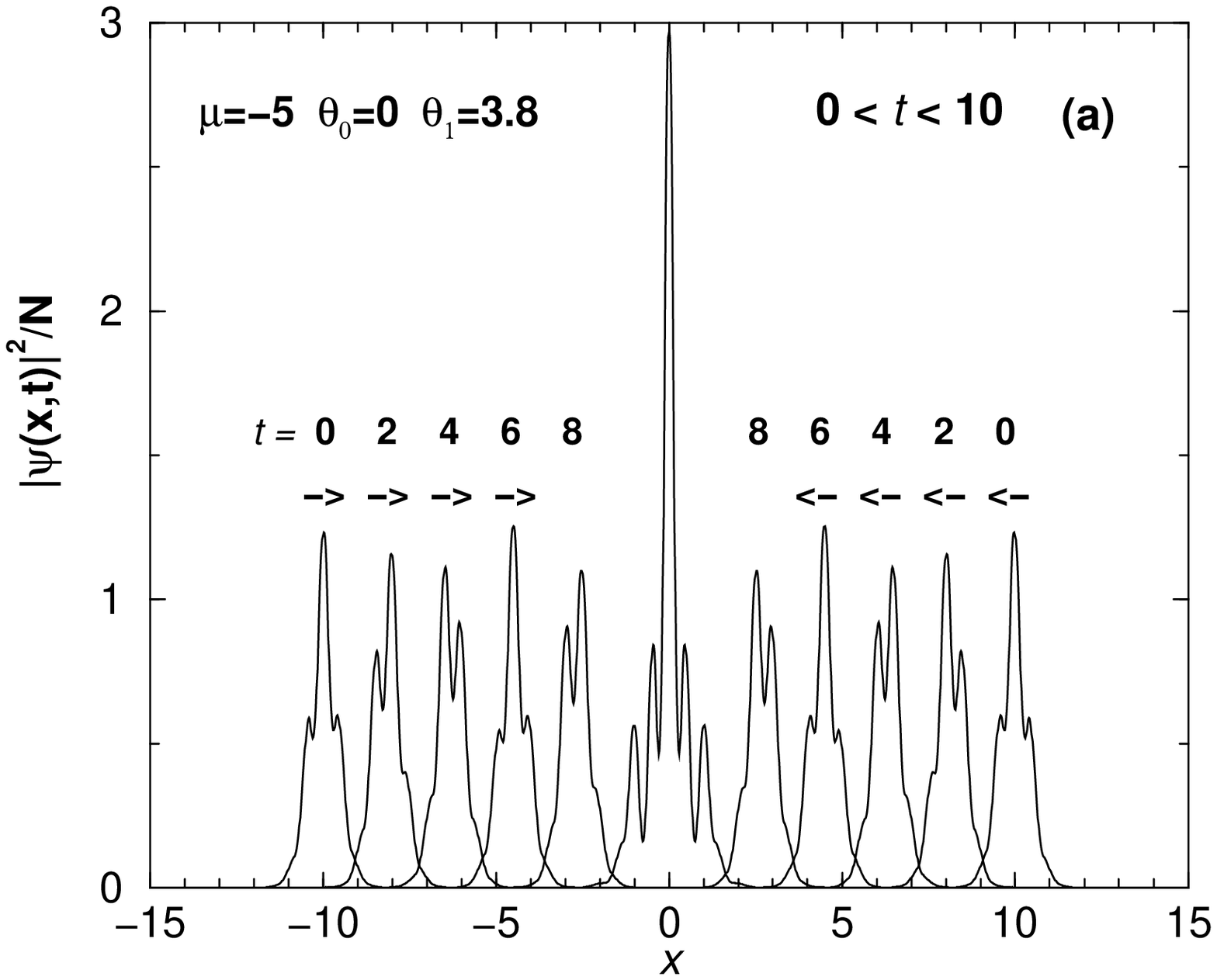}
\hskip 0.2cm
\includegraphics[width=7.5cm,clip]{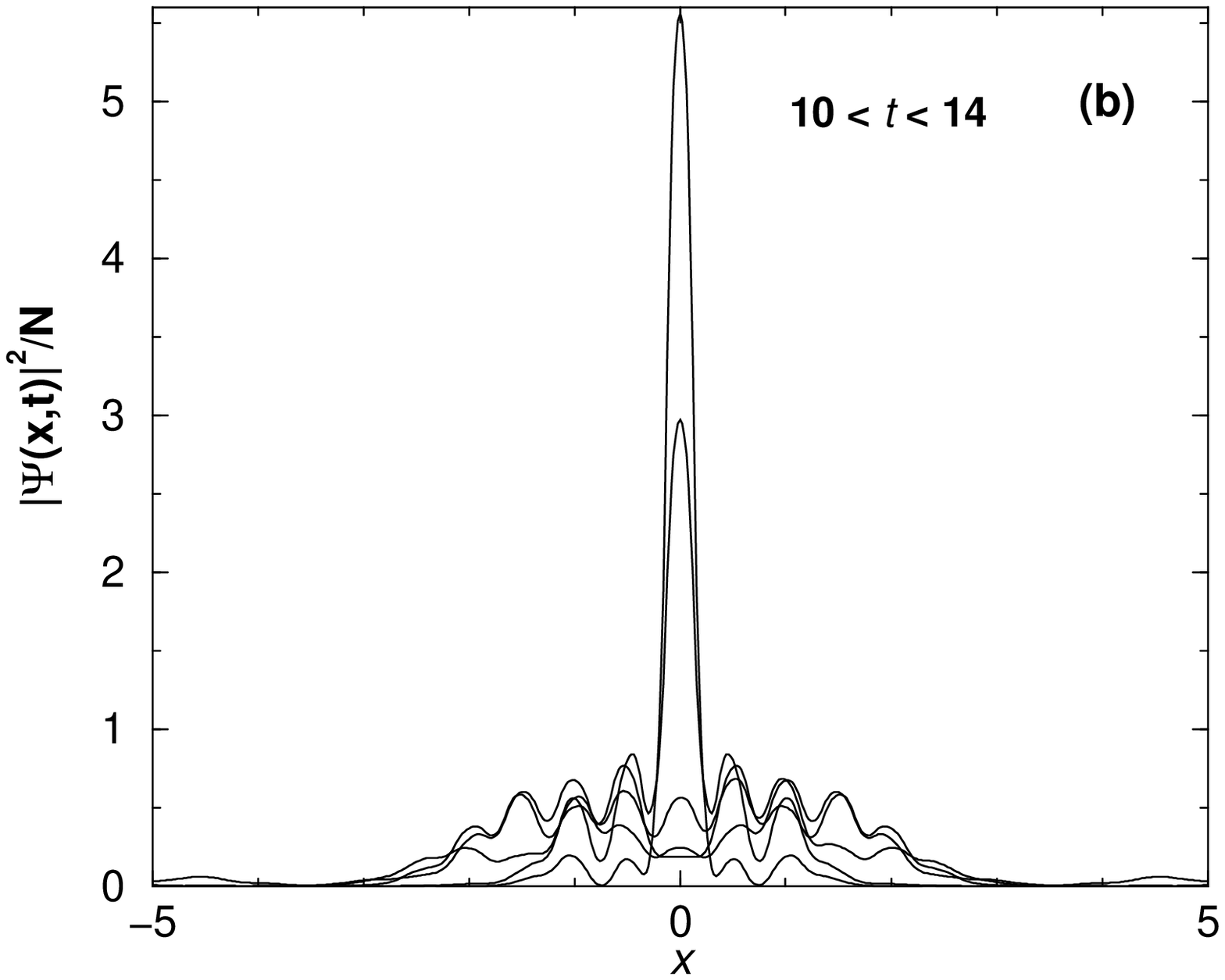}
}
\caption{The inelastic collision of two soliton for $\protect\theta _{0}=0$,
$\protect\theta _{1}=3.8$, with chemical potentials $\protect\mu =-5$ and
initial positions $x=\pm 10$. The panel (a) shows a sequence of
soliton profiles at $0\leq t\leq 10$, through time intervals $\Delta t=2$.
In the panel (b), a sequence of soliton profiles is shown at $10\leq t\leq
14$. The merger of the colliding solitons into a bound state and strong
emission of radiation are clearly identified by the plots. }
\label{Fig-13}
\end{figure}

It is obvious that the present system is far from any integrable limit,
hence inelastic collisions leading to merger of the colliding solitons are
expected too. Two examples of such strongly inelastic collisions are shown
in Figs.~\ref{Fig-12} and \ref{Fig-13}, for the parameters $\theta _{0}=0$
and $\theta _{1}=3.8$. In Fig.~\ref{Fig-12}, the results are plotted for two
solitons with $\mu =-2.5$ and initial velocity $v_{s}=0.3$. The merger of
the solitons into an oscillating localized wave packet is observed here as
the result of the collision, accompanied by emission of radiation. In 
Fig.~\ref{Fig-13}, we present results for the collision of two narrow solitons
with $\mu =-5$. The panels (a) and (b) display the picture before
and after the collision, for time intervals $0\leq t\leq 10$ and $10\leq
t\leq 14$ respectively. The merger into an oscillating localized mode,
surrounded by a cloud of emitted radiation, is clearly seen in this case too.

\section{Conclusion}

In this work, we propose the realization of a nonlinear nonlocal lattice in
quasi-one-dimensional dipolar BECs. The lattice is introduced by a spatially
periodic modulation of the DDI (dipole-dipole interactions), that, in turn,
may be induced by an external field, which periodically changes the local
orientation of permanent dipoles. The necessary setting may be created in
experiments using the currently available MLs (magnetic lattices). We have
focused on the consideration bright solitons, which were constructed in the
numerical form, and by means of the VA (variational approximation). The
stability of the solitons is studied by direct simulations of their
perturbed evolution, and also by means of the VK (Vakhitov-Kolokolov)
criterion~\cite{VK}. The VA provides, overall, a good accuracy in comparison
with numerical results. Stable solitons exist below a critical value of the
angle ($\theta _{0}$) between the average orientation of the dipoles and the
system's axis. The stability area shrinks with the increase of the
modulation amplitude ($\theta _{1}$). Unstable solitons are, typically,
spontaneously transformed into persistent localized breathers (which are not
observed in the system with the uniform DDI).

The mobility of the solitons and collisions between them have been studied
too. Strongly inelastic collisions lead to the merger of the solitons into
oscillatory localized modes, surrounded by conspicuous radiation fields.

The next objective, which will be considered elsewhere, is to study the
model of the condensate composed of polarizable atoms or molecules, in which
the magnitude of the external polarizing field, which induces the local
electric or magnetic moments (in the absence of a permanent moment), varies
periodically along the system's axis. A challenging possibility is to extend
this class of models to two-dimensional settings.

\acknowledgments B.A.M. appreciate a visitor's grant from the South American
Institute for Fundamental Research, and hospitality of Instituto de F\'{\i}
sica Te\'{o}rica at Universidade Estadual Paulista (S\~{a}o Paulo, Brazil).
We also acknowledge support from Brazilian agencies Funda\c{c}\~{a}o de
Amparo \`{a} Pesquisa do Estado de S\~{a}o Paulo and Conselho Nacional de
Desenvolvimento Cient\'{\i}fico e Tecnol\'{o}gico.


\begin{thebibliography}{99}
\bibitem{OL} M. Greiner, O. Mandel, T. Esslinger, T. W. Hansch, and I.
Bloch, Nature \textbf{415}, 39 (2002); D. Jaksch, C. Bruder, J. I. Cirac, C.
W. Gardiner, and P. Zoller, Phys. Rev. Lett. \textbf{81}, 3108 (1998); V. A.
Brazhnyi and V. V. Konotop, Mod. Phys. Lett. B \textbf{18}, 627 (2004); O.
Morsch and M. Oberthaler, Rev. Mod. Phys. \textbf{78}, 179 (2006); M.
Lewenstein, A. Sanpera, V. Ahufinger, B. Damski, A. Sen(De), and U. Sen,
Adv. Phys. \textbf{56}, 243 (2007); Bloch, J. Dalibard, and W. Zwerger, Rev.
Mod. Phys. \textbf{80}, 929 (2008).

\bibitem{Oxford} M. Lewenstein, A. Sanpera, and V. Ahufinger, \textit{\
Ultracold Atoms in Optical Lattices: Simulating quantum many-body systems}
(Oxford University Press:\ Oxford, UK, 2012).

\bibitem{ML} S. Ghanbari, T. D. Kieu, A. Sidorov, and P. Hannaford, J. Phys.
B \textbf{39}, 847 (2006); A. Abdelrahman, P. Hannaford, and K. Alameh, Opt.
Express \textbf{17}, 24358 (2009).

\bibitem{KMT} Y. V. Kartashov, B.A. Malomed and L. Torner, Rev. Mod. Phys.
\textbf{83}, 247 (2011).

\bibitem{Baranov} M. Baranov, Phys. Rep. \textbf{464}, 71 (2008).

\bibitem{Lahaye} T. Lahaye, C. Menotti, L. Santos, M. Lewenstein and T. Pfau
Rep. Prog. Phys. \textbf{72}, 126401 (2005).

\bibitem{Pfau} A. Maluckov, G. Gligori\'{c}, Lj. Had\v{z}ievski, B. A.
Malomed, and T. Pfau, Phys. Rev. Lett. \textbf{108}, 140402 (2012); Phys.
Rev. A \textbf{87}, 023623 (2013).

\bibitem{2D} P. Pedri and L. Santos, Phys. Rev. Lett. 95, 200404 (2005); R.
Nath, P. Pedri, and L. Santos, Phys. Rev. A \textbf{76}, 013606 (2007); I.
Tikhonenkov, B. A. Malomed, and A. Vardi, Phys. Rev. Lett. \textbf{100},
090406 (2008); Phys. Rev. A \textbf{78}, 043614 (2008).

\bibitem{JCuevas} J. Cuevas, B. A. Malomed, P. G. Kevrekidis, and D. J.
Frantzeskakis, Phys. Rev. A \textbf{79}, 053608 (2009).

\bibitem{Luis} L. E. Young-S., P. Muruganandam, and S. K. Adhikari, J. Phys.
B \textbf{44}, 101001 (2011).

\bibitem{KLakomy} K. Lakomy, R. Nath, and L. Santos, Phys. Rev. A \textbf{86}
, 013610 (2012).

\bibitem{Belgrade} G. Gligori\'{c}, A. Maluckov, L. Had\v{z}ievski, and B.
A. Malomed, Phys. Rev. A \textbf{78}, 063615 (2008); \textit{ibid}. \textbf{%
\ 79}, 053609 (2009); J. Phys. B: At. Mol. Opt. Phys. \textbf{42}, 145302
(2009).

\bibitem{SKA} P. Muruganandam, R. K. Kumar, and S. K. Adhikari, J. Phys. B
At. Mol. Opt. Phys. \textbf{43}, 205305 (2010); S. K. Adhikari and P.
Muruganandam, \textit{ibid}. \textbf{44}, 121001 (2011); \textit{ibid}.
\textbf{45}, 045301 (2012); Phys. Lett. A \textbf{376}, 2200 (2012).

\bibitem{RVMA} S. Rojas-Rojas, R. A. Vicencio, M. I. Molina, and F. Kh.
Abdullaev, Phys. Rev. A \textbf{84}, 033621 (2011).

\bibitem{Kartashov} Y. V. Kartashov, V. A. Vysloukh, and L. Torner, Optics
Letters, \textbf{33}, 1774 (2008).

\bibitem{polarizable} S. Yi and L. You, Phys. Rev. A \textbf{61}, 041604
(2000); B. Deb and L. You, Phys. Rev. A \textbf{64}, 022717 (2001); T. J.
McCarthy, M. T. Timko, and D. R. Herschbach, J. Chem. Phys. \textbf{125},
133501 (2006); Z. D. Li, Q. Y. Li, P. B. He, J. Q. Liang, W. M. Liu, and G.
S. Fu, Phys. Rev. A \textbf{81}, 015602 (2010); A. E. Golomedov, G. E.
Astrakharchik, and Yu. E. Lozovik, Phys. Rev. A \textbf{84}, 033615 (2011).

\bibitem{Gershon} S. Giovanazzi, D. O'Dell, and G. Kurizki, Phys. Rev. Lett.
\textbf{88}, 130402 (2002).

\bibitem{Roati} M. Fattori, G. Roati, B. Deissler, C. D'Errico, M. Zaccanti,
M. Jona-Lasinio, L. Santos, M. Inguscio, and G. Modugno, Phys. Rev. Lett.
\textbf{101}, 190405 (2008).

\bibitem{Santos} S. Sinha and L. Santos, Phys. Rev. Lett. \textbf{99},
140406 (2007).

\bibitem{BGT} M. Brtka, A. Gammal, and L. Tomio, Phys. Lett. A \textbf{359},
339 (2006).

\bibitem{YCai} Y. Cai, M. Rosenkranz, Z. Lei, and W. Bao, Phys. Rev. A
\textbf{82},043623 (2010).

\bibitem{BBB} B. B. Baizakov, F. Kh. Abdullaev, B. A. Malomed, and M.
Salerno, J. Phys. B: At. Mol. Opt. Phys. \textbf{42}, 175302 (2009).

\bibitem{VK} M. Vakhitov and A. Kolokolov, Radiophys. Quantum Electron.
\textbf{16}, 783 (1973); L. Berg\'{e}, Phys. Rep. \textbf{303}, 259 (1998).

\bibitem{AGST} F. Kh. Abdullaev, A. Gammal, M. Salerno, and L. Tomio, Phys.
Rev. A \textbf{77}, 023615 (2008).

\bibitem{AGT} F. Kh. Abdullaev, A. Gammal, and L. Tomio, J. Phys. B \textbf{%
\ 37}, 635 (2004); F. Kh. Abdullaev, R. M. Galimzyanov, M. Brtka, and L.
Tomio, Phys. Rev. E \textbf{79}, 056220 (2009).
\end{thebibliography}
\end{document}